\documentclass{article}

 \usepackage[preprint]{neurips_2025}

\usepackage[utf8]{inputenc} 
\usepackage[T1]{fontenc}    
\usepackage{hyperref}       
\usepackage{url}            
\usepackage{booktabs}       
\usepackage{amsfonts}       
\usepackage{nicefrac}       
\usepackage{threeparttable} 
\usepackage{microtype}      
\usepackage{xcolor}         
\usepackage{amsmath}
\usepackage{bm}
\usepackage{graphicx}
\usepackage{subcaption}
\usepackage{float}
\usepackage{pgfplots}
\usepackage{siunitx}
\usepackage{booktabs}
\pgfplotsset{compat=1.18}
\usepackage{tikz}

\usepackage{listings}
\definecolor{codegreen}{rgb}{0,0.6,0}
\definecolor{codegray}{rgb}{0.5,0.5,0.5}
\definecolor{codepurple}{rgb}{0.58,0,0.82}
\definecolor{backcolour}{rgb}{0.95,0.95,0.92}

\lstdefinestyle{mystyle}{
    backgroundcolor=\color{backcolour},   
    commentstyle=\color{codegreen},
    keywordstyle=\color{magenta},
    numberstyle=\tiny\color{codegray},
    stringstyle=\color{codepurple},
    basicstyle=\ttfamily\footnotesize,
    breakatwhitespace=false,         
    breaklines=true,                 
    captionpos=b,                    
    keepspaces=true,                 
    numbers=left,                    
    numbersep=5pt,                  
    showspaces=false,                
    showstringspaces=false,
    showtabs=false,                  
    tabsize=2
}

\title{Fluid Intelligence: A Forward Look on AI Foundation Models in Computational Fluid Dynamics}

\author{%
  Neil Ashton \\
  NVIDIA\\
  Santa Clara\\
  California, US, 95051 \\
  \texttt{nashton@nvidia.com} \\
\And
  Johannes Brandstetter  
  \\
  Emmi AI \\
  Johannes Kepler University (JKU) Linz\\
  Linz, Austria\\
  \texttt{johannes@emmi.ai} \\
  \And
    Siddhartha Mishra \\
  Seminar for Applied Mathematics\\
  D-Math and ETH AI Center, ETH Zurich\\
  Switzerland \\
  \texttt{siddhartha.mishra@sam.math.ethz.ch} \\
}

\begin{document}

\maketitle

\begin{abstract}
Driven by the advancement of GPUs and AI, the field of Computational Fluid Dynamics (CFD) is undergoing significant transformations. This paper bridges the gap between the machine learning and CFD communities by deconstructing industrial-scale CFD simulations into their core components. Our main contribution is to propose the first scaling law that incorporates CFD inputs for both data generation and model training to outline the unique challenges of developing and deploying these next-generation AI models for complex fluid dynamics problems. Using our new scaling law, we establish quantitative estimates for the large-scale limit, distinguishing between regimes where the cost of data generation is the dominant factor in total compute versus where the cost of model training prevails.
We conclude that the incorporation of high-fidelity transient data provides the optimum route to a foundation model. We constrain our theory with concrete numbers, providing the first public estimates on the computational cost and time to build a foundation model for CFD.
\end{abstract}

\newpage

\tableofcontents

\section{Introduction}

Computational Fluid Dynamics (CFD) is omnipresent, influencing nearly every aspect of modern life. From the design of cars \citep{Ashton2018g} and airplanes \citep{clark2025HLPW5}, to the simulation of blood flow in human bodies \citep{Khalafvand2011}, CFD plays a critical role. 
Today, this field is undergoing a disruptive transformation, moving from expert-driven, human-intensive workflows toward a future shaped by AI \citep{lino2023_new,Duraisamy2018}. Historically, the assumption was that increasing High-Performance Computing (HPC) resources would be the primary driver of progress, enabling a gradual transition to higher-fidelity simulations like Large-Eddy Simulation (LES) and Direct Numerical Simulation (DNS). Whilst GPU-accelerated CFD have typically provided an order of magnitude reduction in runtime \citep{Appa2021,nielsen2024,hosseinverdi2025rapidus}, the rise of AI, exemplified by the widespread impact of ChatGPT \citep{achiam2023gpt}, has introduced a new paradigm. This technological shift, largely absent from CFD discourse \citep{Slotnick2014,Spalart2016} until recently, forces a re-evaluation of how we approach complex simulations.

One of the most promising uses of AI within CFD is the development of AI surrogate models. The key advantage of AI surrogate models, trained on prior CFD simulation data, is their ability to make real-time predictions. Generating results in under a second on a modest GPU, they are orders of magnitude faster and more cost-effective than conventional CFD solvers. Whilst state-of-the-art models \citep{alkin2025ab,wu2024transolver,ranade2025domino} are approaching the accuracy of conventional solvers for specific use-cases, their generalizability remains a key question. To date, most industrially relevant publications have utilized small datasets, each tailored to a single use case (e.g., road-car aerodynamics) and a specific CFD simulation setup.

A key question to the CFD and ML communities is whether the development of a foundation model for CFD is possible, just like it has been for general purposes AI (ChatGPT \citep{achiam2023gpt}, Gemini~\citep{team2023gemini}), computer vision~\citep{kirillov2023segment,rombach2022high}, protein folding \citep{jumper_highly_2021,abramson_accurate_2024}, material
design \citep{merchant_scaling_2023,zeni_generative_2025,yang_mattersim_2024}, or
weather/climate (Aurora \citep{bodnar_aurora_2024}).  
A more specific question is ``How can foundation models best be framed for CFD?''. 
This paper aims to answer this question. 

We firstly bridge the gap between the machine learning and CFD communities by deconstructing the CFD process into its fundamental elements and modeling challenges. This part is written as go-to reference for the ML community to understand better the challenges of the CFD community.

The main contribution of this paper is a novel scaling law that incorporates CFD inputs, providing a roadmap for developing and deploying robust foundation models for complex fluid dynamics problems from a dataset and model training perspective. We discuss for which parameters the cost of data generation is the dominant factor in total compute versus where the cost of model training prevails. We conclude that the incorporation of high-fidelity transient data provides the optimum route to a CFD foundation model.
We also pose open questions regarding optimized data generation, online timing, data compression and the co-design of ML and CFD codes. 

\section{Deconstructing the CFD process}

An industry-scale CFD simulation is a three-stage computational pipeline that transforms a high-dimensional problem definition into low-dimensional engineering insights. This pipeline consists of three core elements:

\begin{enumerate}
    \item \textbf{Inputs:} This phase translates a physical problem -- defined by its geometry, physics models, and boundary conditions -- into a discrete mathematical system. The continuous domain is discretized into a computational grid through meshing, and the choice of numerical methods transforms the governing partial differential equations (PDEs) into a large-scale, non-linear system of algebraic equations.

    \item \textbf{Compute engine:} This is the most resource-intensive phase, dedicated to solving the algebraic system. It typically involves iteratively linearizing the problem and repeatedly solving a large, sparse linear system of the form $\bm{A}\bm{x} = \bm{b}$. The choice of algorithm (e.g., Krylov subspace methods, Multigrid) is tightly coupled to the underlying hardware architecture (e.g., CPUs or GPUs).

    \item \textbf{Outputs:} The compute engine yields a high-dimensional raw solution vector. Final quantities of interest; such as aerodynamic forces ($C_L, C_D$) or heat transfer coefficients -- are extracted by applying post-processing operators, often integrals over a surface or volume. This step distills terabyte-scale raw data into the scalar values that drive engineering decisions.
\end{enumerate}

We now detail these steps to highlight the vast input space of a CFD problem, which informs our later discussion on foundation models.

\subsection{Inputs}
\subsubsection{Geometry}

In CFD, a physical object is represented as a bounded domain, $\Omega \subset \mathbb{R}^3$, typically created in a Computer-Aided Design (CAD) package \citep{coons67}. The industry standard for encoding $\Omega$ is the Boundary Representation (B-Rep), which defines the domain by precisely describing its boundary, $\partial\Omega$. This boundary is decomposed into vertices, edges, and faces, whose geometry is most commonly defined using Non-Uniform Rational B-Splines (NURBS) \citep{piegl2012nurbs}.

For design exploration, the geometry can be further parameterized by a vector of $d$ design variables, $\delta \in \mathbb{R}^d$. In this case, the domain becomes a function $\Omega(\delta)$, where changes to $\delta$ (e.g., airfoil chord or pipe radius etc) smoothly deform the geometry by modifying the underlying NURBS control points or weights.

\subsubsection{Geometry pre-processing and clean-up}

A critical stage in industrial CFD is geometry pre-processing. A raw CAD assembly rarely forms a valid computational domain boundary because its boundary, $\partial\Omega$, must be a closed, orientable, ``watertight'' 2-manifold without self-intersections. Raw CAD models frequently violate these requirements with defects like gaps, self-intersections, and non-manifold topology. Resolving these defects is a laborious manual process that can take weeks.

To mitigate this, automated ``shrink-wrapping'' techniques \citep{seo2005} generate a new, approximate surface, $\partial\Omega'$, by projecting a discrete mesh onto the original CAD model. While this yields a usable, closed surface, it introduces geometric error and often adds a serial bottleneck into the computational process. Consequently, moving from a complex CAD model to a high-fidelity CFD simulation without significant manual intervention or geometric approximation remains a major challenge and is often not GPU accelerated. This reliance on an exact boundary representation is a notable difference from standard computer vision tasks, where representations are often discrete (e.g., pixels or voxels) and the goal is often invariant to geometric transformations.

\subsubsection{Physics modeling}

While often framed as solving the Navier-Stokes equations, industrial CFD practice is more complex. For flows at high Reynolds numbers ($Re$), the non-linear convective term leads to turbulence, a chaotic state with a vast range of spatial and temporal scales. Directly resolving all these scales via DNS is computationally prohibitive, with costs scaling as $\mathcal{O}(Re^3)$, thus making the DNS of complex geometries at realistic Reynolds numbers unachievable even with the largest supercomputers available today.  This necessitates turbulence models, which fundamentally alter the system of equations being solved and typically introduce an effective error term between real-life and simulation.

For a Newtonian fluid with constant density $\rho$, dynamic viscosity $\mu$, and external force $\vec{f}$, these equations govern the evolution of the velocity field $\vec{u}(\vec{x}, t)$ and pressure $p(\vec{x}, t)$:
\begin{align}
\rho \left( \frac{\partial \vec{u}}{\partial t} + (\vec{u} \cdot \nabla) \vec{u} \right) &= -\nabla p + \mu \nabla^2 \vec{u} + \vec{f} \\
\nabla \cdot \vec{u} &= 0 \ .
\end{align}

\paragraph{Reynolds-averaged Navier-Stokes (RANS).} The industrial standard, RANS, decomposes flow quantities ($\phi$) into mean ($\bar{\phi}$) and fluctuating components ($\phi'$)

\begin{equation}
    \phi(\vec{x}, t) = \bar{\phi}(\vec{x}) + \phi'(\vec{x}, t) \ .
\end{equation}

This introduces an unclosed term, the Reynolds stress tensor, which is modeled -- often using the Boussinesq hypothesis that relates the Reynolds stress tensor $\overline{u'_i u'_j}$ to the mean strain rate via a turbulent ``eddy'' viscosity $\mu_t$ \citep{Launder_Sandham_2002}.
The full relationship used to model the Reynolds stress tensor is:
\begin{equation}
    -\rho \overline{u'_i u'_j} = \mu_t \left( \frac{\partial \bar{u}_i}{\partial x_j} + \frac{\partial \bar{u}_j}{\partial x_i} \right) - \frac{2}{3} \rho k \delta_{ij} \ ,
\label{eq:rans1}
\end{equation}
where indices $i,j$ indicate the spatial components of $\bar{u}(\vec{x})$. Alternatively, RANS can be closed by directly modeling each of the Reynolds stresses and a length scale quantity~\citep{launder75}. RANS is computationally affordable but its steady-state approach limits accuracy in flows with massive separation \citep{Ashton2016,ashton2023b} or highly anisotropic flows.

\paragraph{Scale-resolving simulations: LES and hybrid methods.} To achieve higher fidelity, LES directly resolves large turbulent eddies and models only the smaller sub-grid scales (SGS). This is achieved by applying a spatial filter (denoted by $\widetilde{\cdot}$ ) to the Navier-Stokes equations, which isolates the unresolved sub-grid scale stress tensor, $\mathcal{T}_{ij}$. Similar to RANS, this unclosed term is most commonly modeled using a Boussinesq-like hypothesis, relating it to the filtered strain-rate tensor $\mathcal{T}_{ij}$:

\begin{equation}
\label{eq:sgs_model}
\mathcal{T}_{ij} \equiv -\rho (\widetilde{u_i u_j} - \tilde{u}_i \tilde{u}_j) = \mu_{sgs} \left( \frac{\partial \tilde{u}_i}{\partial x_j} + \frac{\partial \tilde{u}_j}{\partial x_i} \right) - \frac{2}{3} \rho k_{sgs} \delta_{ij} \ .
\end{equation}

Here, $\mu_{sgs}$ is the SGS (eddy) viscosity and $k_{sgs}$ is the SGS kinetic energy, both of which must themselves be modeled.

A common algebraic closure for the SGS viscosity is the Smagorinsky-Lilly model \citep{smagor1963}. This model links $\mu_{sgs}$ to the local filtered strain rate and the filter size:

\begin{equation}
\label{eq:smagorinsky}
\mu_{sgs} = \rho (C_s \Delta)^2 |\tilde{S}| \ ,
\end{equation}

where:
\begin{itemize}
    \item $C_s$ is the Smagorinsky coefficient, a dimensionless model constant (typically $\approx 0.1 - 0.2$).
    \item $\Delta$ is the filter width, a length scale related to the grid cell size (e.g., $\Delta = (\Delta x \Delta y \Delta z)^{1/3}$).
    \item $|\tilde{S}|$ is the magnitude of the filtered strain-rate tensor, $\tilde{S}_{ij}$.
\end{itemize}

The filtered strain-rate tensor and its magnitude are defined as:
\begin{align}
\label{eq:strain_rate}
\tilde{S}_{ij} &= \frac{1}{2} \left( \frac{\partial \tilde{u}_i}{\partial x_j} + \frac{\partial \tilde{u}_j}{\partial x_i} \right) \\
|\tilde{S}| &= \sqrt{2 \tilde{S}_{ij} \tilde{S}_{ij}} \ .
\end{align}

The cost is orders of magnitude higher than RANS, particularly for Wall-Resolved LES (WRLES), which requires extremely fine near-wall meshes ($y^+ \approx 1$\footnote{The parameter $y^+$ (pronounced ``y-plus'') is a dimensionless distance used to measure the distance of the first computational mesh cell center away from a wall. $y^+ \approx 1$ is the viscous sublayer, where the model must resolve the strong velocity gradients right at the wall, demanding a very fine mesh near the wall. On the other hand, $30 < y^+ < 100$ allows for a wall function approach, where the mesh is coarse near the wall, and the model jumps over the viscous sublayer by applying an empirical law-of-the-wall. This is used in many RANS and Wall-Modeled LES (WMLES) applications.}). To mitigate this, strategies like Wall-Modeled LES (WMLES) \citep{Larsson2015v2} and Hybrid RANS-LES (most commonly the Detached-Eddy simulation (DES) variant \citep{Spalart2009}) are used that reduces the need for spatial resolution (and associated lower time-step choice) in the inner boundary layer. These transient simulations require massive meshes (billions of cells) and small time-steps \citep{cetin2023b}.

The central point is that the engineer's choice of turbulence model (RANS, LES) and any additional physics (e.g., transition, chemistry) results in a different system of PDEs. There is no universal PDE for industrial CFD, a fact with profound implications for developing AI surrogates, which must be conditioned on these modeling choices.

\subsubsection{Meshing}

Meshing discretizes the continuous fluid domain $\Omega$ into a computational mesh. This step is foundational for numerical methods and is tightly coupled with the physics and numerical schemes chosen.

The primary methods in CFD -- Finite Difference (FDM), Finite Volume (FVM), and Finite Element (FEM) -- impose different requirements. FDM typically uses structured grids, which are efficient but struggle with complex geometries. FVM, the dominant approach in commercial codes, uses an integral form of the conservation laws on unstructured cells (e.g., tetrahedra, polyhedra), ensuring conservation and geometric flexibility. FEM offers a rigorous mathematical foundation enabling high-order methods that achieve high accuracy by increasing the polynomial degree of basis functions within elements. However, these methods often require complex, curved meshes to maintain accuracy on curved domains.

Mesh quality, measured by metrics like skewness and orthogonality, directly impacts simulation accuracy and stability. Engineers refine meshes in regions of high gradients (e.g., boundary layers, shock waves), leading to modern simulations using meshes with hundreds of millions to billions of cells. 
Mesh and discretization convergence assures that these errors tend toward zero as the resolution increases. Consequently, the impact of the mesh and the numerical discretization can be removed. However, finer meshes are computationally expensive and often infeasible, thus leading mesh resolution and design to be a key CFD simulation input.
The generation of these massive, high-quality meshes is a significant, often CPU-limited, bottleneck.

\paragraph{Mesh-free and Particle-based methods.}
To circumvent the bottlenecks of grid generation, particularly for complex, moving geometries or free-surface flows, Lagrangian or particle-based methods offer an alternative paradigm. Methods such as Smoothed Particle Hydrodynamics (SPH) \citep{monaghan92} or the Lattice Boltzmann Method (LBM) \citep{chen98} (which, while Eulerian, utilizes a Cartesian lattice that avoids body-fitted meshing) fundamentally alter the input phase.
In these frameworks, the domain $\Omega$ is discretized into a collection of particles or lattice nodes rather than a contiguous mesh. This eliminates the topological constraints of ensuring a watertight, non-intersecting grid, effectively removing the mesh generation bottleneck. However, these methods introduce their own trade-offs, often requiring significantly higher particle counts to resolve boundary layers compared to the stretched cells of a body-fitted RANS mesh.

\subsubsection{Numerical method and discretization}

Linked to meshing is the choice of a numerical framework to solve the governing equations. This requires selecting methods for spatial and temporal discretization. The choice of a spatial scheme for fluxes is crucial for stability and accuracy, ranging from simple central differencing to more robust upwind schemes. Higher-order schemes (e.g., MUSCL, WENO) are used to achieve greater spatial accuracy \citep{BREHM2015184}.
For temporal integration, the semi-discrete system is solved using either:
\begin{itemize}
    \item \textbf{Explicit schemes:} Computationally cheap per step but limited by the strict Courant-Friedrichs-Lewy (CFL) stability condition, often requiring millions of time steps for fine-grid simulations. Well suited for WMLES or WRLES methods.
    \item \textbf{Implicit schemes:} Unconditionally stable, allowing larger time steps, but requiring the solution of a large non-linear system at each step, making them more computationally intensive per step. Well suited for RANS.
\end{itemize}

\subsubsection{Initial and boundary conditions}

A well-posed system requires a complete set of boundary and initial conditions, which also serve as the primary interface for user-driven engineering studies.

\paragraph{Boundary conditions.} These prescribe the solution's behavior on the domain boundary, $\partial\Omega$. Common types include Dirichlet (specifying a variable's value), Neumann (specifying its normal derivative), and Robin (specifying a linear combination of the two). Proper specification is critical for the solution's existence and uniqueness. Practically speaking for any given geometry there may be hundreds or thousands of potential boundary condition variables e.g an aerodynamic database of an aircraft flight envelope.

\paragraph{Initial conditions.} For time-dependent problems, the system's state must be defined at $t=0$. A good initial condition can significantly accelerate convergence, while a poor one can lead to instability. For systems with multiple stable states, the initial condition can determine which solution is found, making it a critical parameter for capturing path-dependent phenomena.

\subsection{Compute engine}

The preceding stages culminate in a large-scale system of coupled, non-linear algebraic equations. Solving this system is the most computationally intensive phase of the CFD workflow. For a steady-state problem, the goal is to find a state vector $\bm{Q}$ that satisfies $\bm{R}(\bm{Q}) = 0$. E.g., for an incompressible flow the state vector $\bm{Q} = [u_x, u_y, u_z, p]^T$ comprises the velocity vector $\vec{u}$ and the scalar pressure $p$. The function $\bm{R}$ takes the current approximation of $\bm{Q}$ as input. If $\bm{Q}$ perfectly satisfies the conservation laws (mass, momentum, and energy) in every cell, then $\bm{R}(\bm{Q})$ is zero. In practice, the solver starts with a poor guess for $\bm{Q}$, and $\bm{R}(\bm{Q})$ is large. The value of $\bm{R}(\bm{Q})$ represents the error or imbalance of the conservative flux across each cell boundary. This non-linear system is typically linearized using a Newton-like method, which requires iteratively solving a large, sparse linear system. Key iterative solvers include:
\begin{itemize}
    \item \textbf{Krylov subspace methods:} Krylov subspace methods are the workhorses of modern CFD, including Conjugate Gradient (CG) for symmetric systems and Generalized Minimal Residual (GMRES) or BiCGSTAB for non-symmetric systems.
    \item \textbf{Multigrid methods (MG):} MG methods achieve optimal $\mathcal{O}(N)$ complexity by accelerating convergence using a hierarchy of coarser grids to efficiently damp low-frequency errors.
\end{itemize}

\subsubsection{Hardware architectures}

The co-design of numerical solvers and hardware architecture is critical for performance. The evolution from latency-optimized CPU architectures to throughput-oriented GPUs necessitates a re-evaluation of algorithmic strategies to maximize computational efficiency \citep{Witherden2013}.

The suitability of a numerical algorithm for a given hardware architecture can be quantified by its arithmetic intensity, $I$, defined as the ratio of floating-point operations (FLOPs) to the bytes of data moved from memory:
\begin{equation*}
    I = \frac{\text{FLOPs}}{\text{Bytes}} \ .
\end{equation*}
CPU architectures, designed for low-latency execution of serial or complex conditional logic, are effective for algorithms with low arithmetic intensity. In contrast, GPU architectures are designed to maximize throughput for highly parallel computations. Their performance is most pronounced on ``compute-bound'' algorithms, where $I$ is high and the primary performance limiter is the raw floating-point capability of the processor. This relationship is often visualized using a Roofline model \citep{williams2009}, which plots attainable performance (GFLOP/s) against arithmetic intensity.

\begin{figure}[h!]
    \centering
     \includegraphics[width=0.8\textwidth]{figures/Example_of_a_naive_Roofline_model.svg.png} 
    \caption{A Roofline model illustrating the performance regimes for compute-bound and memory-bound algorithms on different architectures.}
    \label{fig:roofline}
\end{figure}

This architectural dichotomy directly influences solver selection. Explicit methods on structured grids, characterized by local stencil-based operations, exhibit high data locality and high arithmetic intensity. High-order methods, even on unstructured grids are dominated by S/DGEMM (single or double general matrix-multiplication) operations, which are largely compute bound \citep{Witherden2013,vermeire2017}. These properties map well to the Single Instruction, Multiple Threads (SIMT) execution model of GPUs. This is particularly true for particle-based methods (e.g., SPH/LBM), where the absence of a fixed topology and the reliance on local neighbor searches results in extremely high arithmetic intensity, often allowing these solvers to utilize the full floating-point capability of modern hardware.

Conversely, implicit solvers present a more complex challenge. Methods like the Generalized Minimal Residual Method (GMRES) or Multigrid often involve global reduction operations (e.g., dot products) that can be ``memory-bound'' on GPUs, where the performance is limited by memory bandwidth rather than computational speed. However, modern GPU hardware mitigates this bottleneck through features such as high-bandwidth memory (HBM) and high-speed interconnects (e.g., NVLink). These hardware advancements, combined with the development of communication-avoiding algorithms
that restructure computations to minimize data movement, have made implicit methods increasingly efficient on GPU platforms. 

\subsubsection{Mixed precision}

Modern GPUs incorporate specialized hardware, such as Tensor Cores, that offers substantial performance gains by utilizing mixed-precision arithmetic. There is considerable research at present on how best to combine (e.g., \texttt{FP16} or \texttt{FP32}) with higher precision (\texttt{FP64}) to ensure both numerical stability, accuracy and efficiency \citep{siklósi2025reducedmixedprecisionturbulent}. In general utilizing lower precision technique effectively increases the arithmetic intensity of the algorithm, shifting its position on the Roofline plot from the memory-bound to the compute-bound regime.

In addition, lower precision analysis yields compounding efficiency gains. While traditionally applied to error-resilient components like linear solvers, careful implementation of lower precision arithmetic can yield acceptable accuracy for a broad spectrum of CFD numerics. Reducing the storage requirement per grid element allows for addressing larger computational domains or finer meshes without exceeding hardware memory capacities. This reduction also lowers the demand on memory bandwidth, effectively shifting operational intensity towards faster tiers of the memory hierarchy (e.g., maximizing L1/L2 cache hits versus accessing DRAM). Moreover, this approach aligns with the architectural design of modern GPUs, which typically prioritize single precision (FP32) performance. Since double precision (FP64) throughput is often a fraction of the FP32 peak FLOPS depending on the architecture, avoiding FP64 where possible prevents severe hardware underutilization. In scenarios where high precision is strictly required, techniques such as Dekker's method can be employed to improve accuracy, effectively emulating higher precision on standard hardware to maintain numerical fidelity

The result of this hardware-software synergy is a significant increase in computational capability. For suitable problems, performance gains exceeding an order of magnitude over CPU-based implementations are frequently reported in the literature \citep{turner2021,Appa2021,nielsen2024,hosseinverdi2025rapidus}. This level of acceleration makes high-fidelity simulations, such as WMLES, computationally tractable for routine engineering analysis.

\subsection{Outputs}

The final stage is post-processing, where the solver's raw, high-dimensional output is analyzed to extract key engineering metrics.
\begin{itemize}
    \item \textbf{Integrated quantities:} Metrics like lift ($C_L$) and drag ($C_D$) coefficients are calculated by integrating pressure and viscous forces over a surface.
    \item \textbf{Field data:} Analysis often focuses on boundary data like surface pressure distribution or 2D cross-sections. 3D flow structures are visualized with iso-surfaces and streamlines.
    \item \textbf{Higher-order statistics:} For scale-resolving simulations like LES, statistics like velocity derivatives and turbulent fluctuations are captured.
\end{itemize}

High-fidelity methods like WMLES on billion-cell grids create a severe data bottleneck, as saving thousands of multi-gigabyte time steps is impractical. The field is increasingly adopting \textit{in-situ} visualization, a co-processing paradigm where analysis and rendering occurs concurrently with the simulation, operating directly on data in memory. Only the much smaller, post-processed results are saved, enabling analysis of petascale simulations and inspiring the online training of AI models which is discussed later in the paper.

\subsubsection{Steady-state, time-averaged, and instantaneous quantities}

A critical distinction for generating AI training data is the nature of the CFD solution:
\begin{itemize}
    \item A steady-state solution, $u(\vec{x})$, is the converged, time-invariant goal of a typical RANS simulation.
    \item An instantaneous solution, $u(\vec{x},t)$, is the direct output of a time-accurate simulation like LES.
    \item Time-averaged solutions, $\langle u \rangle (\vec{x})$, are generated for scale-resolving simulations by integrating instantaneous data over time, yielding mean performance metrics analogous to steady-state but with the accuracy from the underlying LES-like methods. 
\end{itemize}

The majority of AI surrogate developments have typically only trained on time-averaged or steady-state data. This topic will be major focus of our later discussion.

\subsection{Summary}
In the preceding sections, we've systematically detailed the extensive inputs required for a single CFD simulation, spanning from the initial geometric definition and meshing, through the choices of physics models and numerical methods, to the specification of boundary and initial conditions. Each of these elements represents a dimension in the vast conditional space that governs the simulation's outcome. This detailed overview serves to highlight the sheer scale of the challenge and informs our subsequent discussion on how foundation models can be structured to navigate this high-dimensional input space and learn the intricate mapping from problem definition to engineering insight.

\section{AI surrogate modeling for CFD}

AI represents a new disruptive technology that has potential use cases across the input, compute engine and output phases of a typical CFD simulation. It can be used in:

\begin{itemize}
    \item Development of improved turbulence, transition or chemistry models that plug within a conventional CFD solver
    \item Initialization of CFD solutions to speed up the time to reach a converged or time-averaged solution using an AI surrogate model prediction
    \item Extraction and/or analysis of flow-features using vision models. 
    \item Super-resolution tasks e.g map from RANS to LES fields or coarse to fine meshes
    \item AI surrogate models to replace conventional solvers and predict CFD outputs
    from an input geometry and/or boundary and initial conditions
\end{itemize}

Whilst all have the potential to improve upon existing CFD methods (see Appendix \ref{sec:appendixA} for further discussion), the use-case with the greatest potential for runtime and cost savings are AI surrogate models. When these surrogates are trained with a broad enough training dataset, they are frequently referred to foundation models for CFD \footnote{Note that there is broader discussion around foundation models for weather/climate, physics and science in general.}. This paper is dedicated to helping frame the discussion around the development of these models and to provide open key questions to the community.       

\subsection{AI fundamentals} 
More traditional machine learning (ML) paradigms, particularly supervised learning tasks, are centered on the objective of learning a conditional probability distribution, 
$P(w|v)$,
where $v$ and $w$ are sampled from random input and output variables, $V$ and $W$, respectively. 
The objective is to minimize a loss function $\mathcal{L}$ over a fixed dataset $\mathcal{D} = \{ v_i, w_i \}_{i=1}^N$, such that
\begin{equation*}
    \underset{\theta}{\min}\mathbb{E}_{(v,w)\sim \mathcal{D}} \left[ \mathcal{L}(f_{\theta})(v,w)\right] \ ,
\end{equation*}
where $f_{\theta}$ is the model parameterized by parameters $\theta$. In stark contrast, foundation models represent a paradigm shift, operating on a fundamentally different principle. Instead of being trained to learn a specific conditional distribution, foundation models are designed to learn a joint probability distribution, $P(v,w)$, across a vast and diverse corpus of data that may span multiple modalities, including e.g., text, images, audio, and video. This approach allows the model to develop a deep, generalized understanding of the underlying relationships and structures inherent in the data. The objective is to learn a latent representation of the world, rather than a task-specific mapping, mathematically expressed by modeling the likelihood of the entire dataset, often through a generative process
\begin{equation*}
\underset{\theta}{\max}\mathbb{E}_{z\sim\mathcal{D}}\left[ \log P_{\theta}(z) \right] \ ,
\end{equation*}
where $z$ is a sample from the joint distribution of the training data. This approach allows the model to capture the complex, multi-modal relationships within the data, leading to a latent representation that can be fine-tuned for many different tasks.

\paragraph{Adaption of foundation models to downstream tasks.}
The training of a foundation model on a vast, multi-modal corpus enables it to learn a deep, generalized understanding of the relationships between different data types. 
The goal of adapting a pre-trained foundation model for a specific downstream task, is to leverage its pre-existing knowledge of the joint distribution $P(v,w)$. This adaptation can be mathematically formulated in two ways:

\begin{enumerate}
    \item \textbf{Fine-tuning}, where some of the model's parameters are updated using a small, task-specific dataset, $\mathcal{D}_{\text{task}} = \{(v_i, w_i)\}_{i=1}^M$, with $M\ll N$ used for pre-training. The objective is to slightly shift the model's learned joint distribution to approximate a new conditional distribution, $P(w|v, \text{task})$. This can be seen as minimizing a loss function $\mathcal{L}_{\text{task}}$:
    \begin{equation*}
        \min_{\theta'} \mathbb{E}_{(v,w) \sim \mathcal{D}_{\text{task}}} \left[ \mathcal{L}_{\text{task}}(f_{\theta'}(v), w) \right] \ ,
    \end{equation*} 
    where $f_{\theta'}$ is the fine-tuned model, and $\theta'$ is an update of the original parameters $\theta$ (i.e., $\theta' = \theta + \Delta\theta$).

    \item \textbf{In-context learning}, where the model uses its pre-trained, static joint distribution to infer the correct output for a given input without any change to its parameters. For a given user-specified input, $v_{\text{prompt}}$, the model uses its learned distribution to generate a response $w_{\text{response}}$ that is probabilistically consistent with the prompt's context. This can be viewed as an inference process:
    $$ w_{\text{response}} = \underset{w}{\arg\max} \, P(w | v_{\text{prompt}}) \ . $$
    Here, the model's internal representation of the joint distribution $P(v,w)$ is used to condition the output on the new input, demonstrating its zero-shot or few-shot learning capability. \textbf{Reasoning} extends in-context learning by enabling the model to generate a step-by-step explicit thought process to solve a complex problem. The model uses its learned joint distribution to model the intermediate steps required for a logical solution, often presented as a ``chain-of-thought''. 
\end{enumerate}

For drawing analogies to CFD workflows, we will mostly use the concept of fine-tuning during this paper.

\subsection{Conditional and joint distributions when modeling CFD}
As previously discussed AI surrogates offer the biggest gain in terms of runtime and cost over conventional CFD because the inference of a pretrained model to a new flow condition or geometry is typically less than a second (compared to many hours with conventional simulations) and can often take a geometry or surface mesh as input -- bypassing the time-consuming volume mesh generation stage. 

When modeling the conditional distribution $P(w_{\text{CFD}}|v_{\text{CFD}})$, it is important to take into account the fact that $v_{\text{CFD}}$ is a composite parameter vector that depends on several choices \footnote{We note that this is not an exhaustive list of potential inputs} within the workflow:
\begin{equation} 
\label{eq:x_cfd}
v_{\text{CFD}} = [v_{\text{geom}},
v_{\text{preprocess}},
v_{\text{boundary/initial}},
v_{\text{mesh}}, v_{\text{phys}}, v_{\text{discr}}] \ ,
\end{equation}
where $v_{\text{geom}}$ represents geometry, $v_{\text{preprocess}}$ the geometry preprocessing, $v_{\text{boundary/initial}}$ represents the boundary/initial conditions, $v_{\text{mesh}}$ the meshing topology and resolution, $v_{\text{phys}}$ the physics modeling (e.g., turbulence model), and $v_{\text{discr}}$ the numerical discretization. Each of these choices significantly influences the resulting output distribution. 
$v_{\text{geom}},
v_{\text{preprocess}},
v_{\text{boundary/initial}},
v_{\text{mesh}}$ are continuum variables, whereas $v_{\text{phys}}, v_{\text{discr}}$ are categorical variables, i.e., labels on the choices of the turbulence model and on the discretization (e.g., central vs upwind scheme).

For example two of the currently most widely used datasets for automotive aerodynamics CFD are the DrivAerML~\cite{ashton2024drivaer} and DrivAerNet(++)~\cite{elrefaie2024drivaernet} datasets. On the face of it, they both take an open-source road-car geometry; DrivAer \citep{Heft2014} \footnote{The DrivAerNet++ contains three variants of the DrivAer; notchback, fastback and estate but we restrict our comparison here to just the notchback which they both have in common} and use CFD to simulate the flow over each, maintaining a constant CFD simulation setup for every geometry variation ($\bm{x}_{\text{geom}}$). However if we look closer we can see that each of the CFD inputs $v_{\text{preprocess}}$,$v_{\text{phys}}$,$v_{\text{discr}}$,$v_{\text{mesh}}$, $v_{\text{boundary/initial}}$ are different, as summarized shown in Table \ref{tab:cfd_results}. 

\begin{table}[h]
  \caption{CFD input space across two datasets.}
  \label{tab:my_results_vertical}
  \centering
  \small
  \label{tab:cfd_results}
  \begin{tabular}{l l l}
    \toprule
     & \textbf{DrivAerML} & \textbf{DrivAerNet++} \\
    \midrule
    Total simulations & 487 & 8000 \\
    $v_{\text{geom}}$ & Notchback & [Notchback, estate, fastback] \\
    $v_{\text{preprocess}}$ & Manual ansa & Manual ansa \\
    $v_{\text{phys}}$ & Transient HRLES & Steady-state RANS \\
    $v_{\text{mesh - vol (surf)}}$ & 160M (9M) & 23M (2M) \\
    $v_{\text{boundary/initial}}$ & Static wheels & Rotating wheels \\
    $v_{\text{discr}}$ & Central/upwind & Upwind \\
    \bottomrule
  \end{tabular}
\end{table}

These distinct choices result in two vastly different conditional output distributions and reflect a broader question; what contributes to the optimum dataset composition? Is it dataset size, diversity of geometry/boundary conditions, or the fidelity of the simulation approach (i.e., sim-real delta)?

\begin{figure}[h!]
    \centering
     \includegraphics[width=0.8\textwidth]{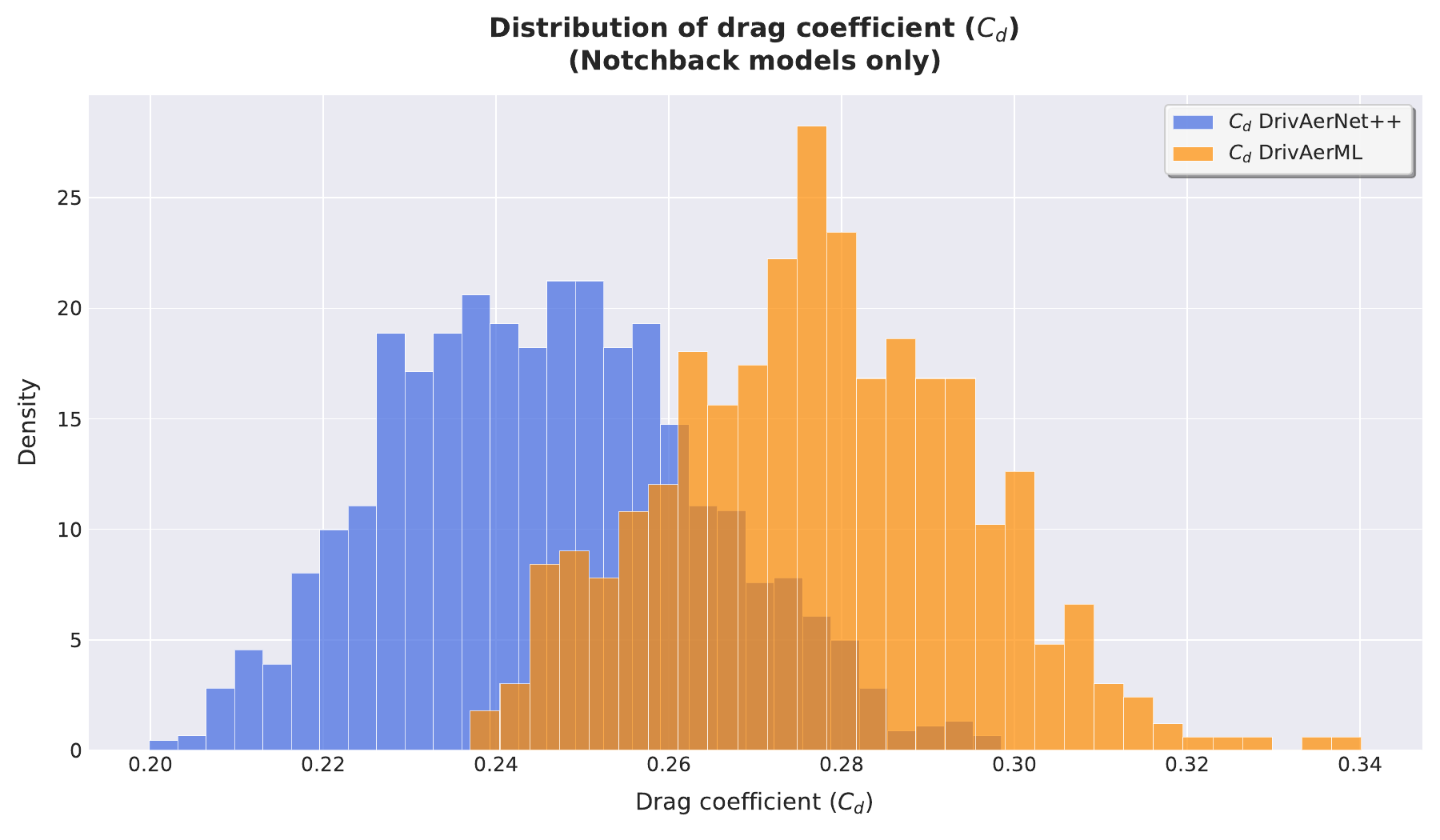} 
    \caption{Drag coefficient distribution for the DrivAerML and DrivAerNET++ datasets. Dark yellow areas represent points of overlapping drag coefficient.}
    \label{fig:drivaerdrag}
\end{figure}

As shown in Figure \ref{fig:drivaerdrag},  these different CFD inputs lead to outputs with limited overlap, making a single conditional model inadequate to capture both distinct, but partially disjoint distributions. Simply put, training a model on both datasets may not ultimately yield a more functional/generalizable model for the prediction of a DrivAer-like body. 

\subsection{Additional data beyond CFD}
The question of CFD fidelity and the sim-real gap is important as CFD itself is a numerical approximation of the Navier-Stokes equations that in turn represent the real-world nature of fluid movements for a given set of initial and boundary conditions. Thus, the conditional distribution of Equation~\eqref{eq:x_cfd}
can be reduced to 
$$ v_{\text{exp}} = [v_{\text{geom}},
v_{\text{boundary/initial}},v_{\text{exp-setup}}] \ .$$

This can be explained by the $v_{\text{mesh}}, v_{\text{preprocess}}, v_{\text{phys}},v_{\text{discr}}$ all being CFD related inputs that are removed when we consider experimental setups where there is no modeling of the physics. An additional term does arise which is the error related to the exp-real gap i.e the influence of the wind-tunnel walls, the way the object is mounted in the wind-tunnel etc. 

Thus, one could conclude that to reduce the CFD input space one should focus on experimental data. However whereas a CFD simulation produces a continuous solution over the entire spatial domain $\Omega \in \mathbb{R}^3$, with outputs
$v_{\text{CFD}}(\vec{x})$ for spatial positions $\vec{x} \in \Omega$, additional data from real-world measures is challenging for a number of reasons. Firstly, physical tests (i.e., flight tests, road-tests, in-situ data) are typically sparse in data, meaning only data is available at specific points in the flow domain. Consequently, real-world physical tests provide data at a discrete and finite set of $N_P$ measurement points $\displaystyle \mathcal{P} = \{ m_i \}_{i=1}^{N_P}$,  and the data is often constrained to the surface $\partial \Omega$ or specific 2D planes using techniques such as particle image velocimetry (PIV)~\citep{raffel2018particle,adrian2011particle}. Additionally, there are typically areas which are hidden from view, i.e., underneath a car. 

Secondly for real-world tests (i.e not wind-tunnels) a major challenge is the variability of flow conditions. 
Real-world tests introduce variability that is difficult to control. This variability means that two tests run under seemingly identical conditions, $v_\text{test}$, will not produce the exact same results. Assuming that the flow condition $v_\text{flow}$ is drawn from a probability distribution $v_\text{flow} \sim P(v_\text{boundary/initial})$, i.e.,
\begin{equation*}
P(w|v) = \mathbb{E}_{v_\text{flow} \sim P(v_\text{boundary/initial})} \left[ P(w | v_\text{geom}, v_{\text{flow}}) \right] \ ,
\end{equation*}
an ensemble of tests would be required given the nature of flow speeds, cross-winds and errors in measurement equipment itself.

\subsection{AI surrogate modeling for CFD}

Generally speaking, a surrogate~\citep{forrester2008engineering} (also known as a surrogate model or emulator) is a model that approximates the behavior of a more complex or computationally expensive system. Surrogate models act as a stand-in for simulations, physical experiments, or mathematical models that are too slow, costly, or difficult to run repeatedly. 

As discussed previously a major challenge is that a single AI surrogate model is often not general enough to capture the different distributions that arise from distinct choices within the CFD pipeline. 
 
Mathematically, training an AI surrogate involves learning a function $\hat{\mathcal{G}}$ that approximates the true solution operator $\mathcal{G}$ of either the governing PDE, or the input-output mapping specified by user choices (e.g., meshing or discretization). The core objective is thus to find a model $\hat{\mathcal{G}}$ such that the conditional distribution $P(w_\text{CFD}|v_\text{CFD})$ is approximated:
\begin{equation*}
P(w_\text{CFD}|v_\text{CFD}) \approx P(\hat{\mathcal{G}}(v_\text{CFD})) \ ,
\end{equation*}
where any combination of the composite parameter vector $v_{\text{CFD}}$ according to Equation~\eqref{eq:x_cfd} can be input the surrogate. For example, in most current CFD datasets, the turbulence model (part of $v_{\text{phys}}$), the meshing procedure $v_{\text{mesh}}$ and the discretization $v_{\text{discr}}$ are fixed.

\subsubsection{Primer on current surrogate modeling paradigms.}
The first wave of AI surrogates for CFD were based on architectures like Graph Neural Networks (GNNs)~\citep{kipfgcn,pfaff2020learning}, Convolutional Neural Networks (CNNs) (e.g., UNet)~\citep{ronneberger2015u,gupta2022towards,CNO,tompson2022acceleratingeulerianfluidsimulation,guo2016}, and Fourier Neural Operators (FNOs)~\citep{Li:20,Li:23}. Many of these models are realizations of the neural operator learning paradigm~\citep{Lu:19, Lu:21, Li:20graph, Li:20,Kovachki:21,bartolucci2023representation,alkin2024universal,ranade2025domino}, where the network learns a mapping between infinite-dimensional input and output function spaces, i.e., $P(\bm{v}_\text{CFD}|\bm{u}_\text{CFD})$. 
Recently, there has been a surge in transformer-based~\citep{vaswani2017attention} surrogates, reflecting their success in other scientific domains such as protein folding~\citep{jumper_highly_2021,abramson_accurate_2024} and weather modeling~\citep{bi_accurate_2023,bodnar_aurora_2024}. Examples of these models include OFormer~\citep{li2022transformer}, Transolver~\citep{wu2024transolver,luo2025transolver++}, AB-UPT~\citep{alkin2025ab}, scOT \cite{herde2024poseidon} and GAOT \cite{wen2025geometry}. These architectures extend the neural operator paradigm by leveraging attention mechanisms to effectively integrate information across different spatial locations and scales, using self-attention, cross-attention, or perceiver blocks~\citep{jaegle2021perceiver,jaegle2021perceiverIO}.

Physics-Informed Neural Networks (PINNs)~\citep{raissi2019physics} represent a different learning paradigm, integrating the governing physical laws of a system directly into the neural network’s learning objective.
This is achieved by including a physics-based term in the loss function, which is minimized
during training via backpropagation with respect to the input coordinates. So far, PINNs have not been successfully applied to complex large-scale problems such as external aerodynamics.

The primary input for an AI surrogate is the geometry of the physical system, representing the object in the computational domain, often represented as a high-fidelity volumetric mesh and the boundary conditions. Mathematically, this mesh represents a discrete sampling of the domain, $\Omega \in \mathbb{R}^3$, and its boundary, $\partial \Omega$. The input to the neural network can be conceptualized as a set of $N_P$ discrete data points, $\mathcal{P} = \{ \vec{x}_i \}_{i=1}^{N_P}$, where each point $\vec{x}_i$ is a vertex in the mesh and is associated with a specific feature vector. A more advanced input approach could be the use of raw computer-aided design (CAD) geometry. This bypasses the need for the computationally expensive meshing process and could be handled by a network that operates on a point cloud or a signed distance function representation of the geometry.
The outputs of AI surrogates are the predicted fields and integral quantities of the fluid flow: (i) typical fields such as pressure or velocity fields  are scalar and vector fields, respectively, defined over the boundary of the geometry or in the 3D volume; (ii) important engineering outputs are the integral quantities derived from the pressure and viscous stress fields over the object's surface. 

Generalization of AI surrogates can occur across different axes of the composite parameter vector $v_{\text{CFD}}$ of Equation~\eqref{eq:x_cfd}, e.g., across geometry variants, inflow conditions, meshing schemes, etc.

\section{From LLM scaling laws to CFD, analogies and differences}

Before moving to the topic of foundation models for CFD, we will first discuss the broader development of LLMs, that we will later use to compare and contrast to similarities and differences in the context of developing such models for CFD.

\subsection{Motivation: Empirical power laws for LLMs}

The modern understanding of AI scaling is largely built upon the seminal work on Transformer-based large language models (LLMs)~\citep{vaswani2017attention,radford2018improving}. Research has empirically demonstrated that the performance of these models, typically measured by the cross-entropy loss ($L$) \footnote{The idea is that models with a lower loss are likely more capable and perform better on downstream tasks as well.} on a held-out test set, scales as a predictable power law with three primary factors: the model size ($M$), defined as the number of non-embedding trainable parameters; the dataset size ($D$), measured in tokens; and the total training compute ($C$)~\citep{kaplan2020scaling,hoffmann2022training}.

This relationship, which holds across many orders of magnitude, can be mathematically formulated as an additive model where the total loss is the sum of an irreducible loss floor and error terms that decay with model and dataset size:

\begin{equation}
L(N, D) \approx \left( \frac{A}{M^\alpha} \right) + \left( \frac{B}{D^\beta} \right) + L_0
\end{equation}

In this formulation, $A$, $B$, $\alpha$, and $\beta$ are constants determined empirically for a given model family and data distribution. The term $L_0$ represents the irreducible loss, an asymptotic performance limit corresponding to the entropy of the true data distribution. This law's predictive power has been a cornerstone of modern AI development, establishing a clear strategy: to achieve breakthrough performance, one must scale up model size, dataset size, and compute in tandem~\citep{hoffmann2022training}. 

Beyond text, the scaling law methodology was successfully applied and validated in computer vision~\citep{zhai2022scaling,dehghani2023scaling}, leveraging the properties of Vision Transformers (ViT)~\citep{dosovitskiy2020image}. 
Modern LLMs such as Gemini~\citep{team2023gemini} or GPT~\citep{achiam2023gpt} have evolved beyond text-only capabilities and are now inherently multi-modal. 

\paragraph{The ``20:1 rule of thumb''.}
\citet{kaplan2020scaling} found that it seemed to be more effective to allocate a larger portion of the increased compute budget to growing the model size rather than the dataset size. This leads to a ``bigger is better'' paradigm.
However, this was challenged and widely replaced by the Chinchilla scaling laws~\citep{hoffmann2022training} where in contrast to~\citet{kaplan2020scaling} the core assumption is that for compute-optimal training, the model size and the number of training tokens should be scaled proportionally and in a balanced manner.  This led to the now-famous ``20:1 rule of thumb'', which suggests that for a given compute budget, a model's parameter count should be roughly proportional to its training data size, specifically requiring about 20 training tokens for every model parameter.
Scaling the number of parameters happens by two processes: (i) by increasing the number of layers allows which the model to learn more complex, hierarchical representations of the data; (ii) increasing the hidden dimension (the size of the internal representations) which allows the model to capture richer features within each layer. 

\paragraph{The ``All Data is Seen Only Once'' assumption.}
A key, often unstated, assumption in the Chinchilla scaling laws and much of the related research is the idea of training for one epoch, or seeing all the training data only once. This assumption is crucial because it ensures that performance gains are coming from the scale of the model and data, not from repeatedly showing the model the same information. In practice, models are often trained for multiple epochs, especially when data is limited, but the ideal scaling laws are based on this single-pass scenario to prevent the model from simply memorizing the training data. This assumption also ties into the concept of compute-optimal training: if you have a fixed compute budget, it's more efficient to train a smaller model on more data (and thus for more time) than to train a massive model on a small dataset for a shorter time.

\paragraph{The analogy for documents and tokens in CFD.}
The training of Large Language Models (LLMs) relies on extensive document corpora (e.g., RedPajama~\citep{weber2024redpajama}, the FineWeb datasets~\citep{penedo2024fineweb}, or DataComp-LM~\citep{li2024datacomp}), which are tokenized into sequences for training. A crucial finding in this domain is that document diversity is a key determinant of a model's performance~\citep{touvron2023llama}. Analogously, similar scaling principles should emerge for AI models for Computational Fluid Dynamics (CFD). In this context, e.g., input variation (see Equation~\eqref{eq:x_cfd}) can be viewed as a ``document'', with different subsets of the simulation mesh serving as training sequences. This approach mirrors the LLM scaling law analogy with two primary tenets: (i) each subset of the simulation mesh is processed only once, ensuring data efficiency (e.g., each DrivAerML dataset geometry can be easily cut in 1000 different subsets with 100k mesh points each, which is sufficient to train proper surrogates~\citep{alkin2025ab,Li:23}); and (ii) input diversity is a primary factor in fostering model performance, directly paralleling the role of document diversity in LLMs.

\paragraph{The analogy for context in CFD.}
Transformer-based LLMs operate fundamentally by processing a sequence of input tokens to predict the next most probable token. This process is critically dependent on the model's context, defined as the full sequence of tokens provided as input. The context acts as the model's working memory, providing a rich, dynamic dataset from which to infer patterns, facts, and stylistic cues. At inference time, the quality and content of the prompt \footnote{Note that the context and prompt are not the same thing.} directly influence the model's performance. A well-known heuristic for effective prompting is to first provide the model with general, domain-specific information before asking a specific question. This method, often referred to as ``priming'' enriches the context, allowing the LLM to access and leverage relevant knowledge embedded within its vast parameter space. This principle of context-dependent performance is directly mirrored in advanced pretraining strategies. Modern LLMs are not trained with a single, fixed context length. Instead, they often undergo a staged pretraining process where the context window is progressively increased~\citep{grattafiori2024llama}.

It is however important to note that LLMs do not receive explicit metadata about the source of the documents they process (e.g., whether the text originates from Wikipedia, Reddit, or a scientific journal). Instead, the model's understanding of a document's source and style is derived implicitly from the contextual cues present in the text itself. 

In stark contrast to LLMs, the training task for AI surrogates for CFD is fundamentally different. Unlike LLMs, CFD surrogates require an explicit parameter vector to define the properties of the underlying simulation. While geometric variations can be encoded through the input mesh, crucial simulation parameters such as preprocessing schemes, boundary and initial conditions, meshing procedures, physics models (e.g., RANS vs. LES), and numerical discretizations cannot be implicitly inferred. For these models, information that sets the scene for a given data point -- analogous to an LLM's understanding of a document's source from its context -- must be provided explicitly to ensure accurate training and generalization across multiple datasets.

\subsection{CFD is not language}
The above discussion clearly indicates that many aspects of LLM training and inference, particularly scaling laws, may be \emph{transferred} to foundation models for CFD. However, CFD is not \emph{natural language} as there are fundamental differences between the two domains. 

To begin with, all LLMs rely on the paradigm of autoregressive \emph{next token prediction}, given an input sequence of tokens (context window), notwithstanding recent developments in discrete diffusion \citep{nie2025largelanguagediffusionmodels}. These tokens are the outputs of a tokenization process that converts text (images, videos etc in multi-modal models) into a set of \emph{discrete, quantized} entries in a codebook~\citep{sennrich2016neural}. During training, the objective is to minimize the \emph{cross-entropy loss} between the distribution of the predicted tokens and the ground truth token distribution. All the scaling laws for LLMs are based on the scaling of this specific loss function, defined over distributions of quantized tokens. 

On the other hand, all the state-of-the-art AI models for CFD and for physics, in general, follow a very different paradigm. The inputs to the models are \emph{continuous objects} (functions/fields over space and time) as denoted by the input vector in Equation~\eqref{eq:x_cfd} and the corresponding outputs are also continuous functions. The tokenization process in this setting converts these continuous inputs into continuous tokens in suitable latent spaces, for instance through embedding patch tokens \cite{dosovitskiy2020image}. The training aims to minimize mismatches between the model predictions and the ground truth outputs, as measured by continuous norms such as root mean square 
error (RMSE) or mean absolute error (MAE). This framework is exemplified by the growing class of Physics Foundation Models \cite{mpp,dpot,herde2024poseidon,walrus} etc. However, these models almost exclusively focus on inputs and outputs on Cartesian grids and are not suited for industrial scale 3D CFD learning tasks. 

Given this fundamental divergence of paradigms between LLMs and CFD, it is only natural to investigate whether scaling laws, analogous to the ones derived for LLMs, also hold in the setting of foundation models for CFD? And secondly, what are the similarities and differences between the derived scaling laws? We explore these questions below. 

\section{A new scaling law for CFD foundation models}
\label{sec:smscl}
Our aim in this section is to demonstrate how scaling laws for a potential CFD foundation model impact data generation and model training. To this end, we start with Equation~\eqref{eq:x_cfd}, where the inputs for any CFD process are described. These include not just continuum inputs such as initial/boundary conditions, domain geometry etc but also categorical variables, i.e., the underlying physics (turbulence) model and the mesh discretization/numerical method etc. To model a representative configuration, we will fix the categorical variables and vary the input $v_{\text{CFD}}$ only over the continuum variables and for notational simplicity, we denote the resulting input variable by $v$.  Then, the input-to-output ground truth map is given by $S: v \mapsto w$. Here, $w$ refers to a steady-state or (long) time-averaged realization of the underlying flow field. This map corresponds to either a full-scale resolving DNS or experimental data and our objective is to approximate/learn this map. To do this, we have to specify the categorical variables. To cover representative cases, we consider the following two sets of variables: underlying turbulence model, and resulting mesh discretization/numerical method. 

\subsection{Low-fidelity RANS simulator} In this case, we assume that a steady-state RANS type turbulence model is used for simulating the physics. We denote this CFD model by $S_r$ and assume that 
\begin{equation}
\label{eq:rans}
\|S - S_r \| \sim \epsilon_r \ .
\end{equation}
Here, $\|.\|$ refers to an operator norm, taken in expectation (average) over the entire population of instances (samples) of fluid flow and $\sim$ denotes equality up to a constant. Note that $\epsilon_r$ is the smallest (minimum) error that the model $S_r$ can yield. 

\paragraph{Cost of data generation} 
We need to discretize the governing RANS equations with a suitable numerical method of sufficient mesh resolution (mesh converged numerical simulation) to achieve errors $\sim \epsilon_r$. To this end, we fix a mesh size $\Delta_r$.  The resulting number of spatial cells or volumes is denoted by $V_r \sim \Delta_r^{-3}$ \footnote{It is noted that industrial meshes will not have uniform resolution in all three spatial dimensions.}. We assume that the numerical model approximating $S_r$ scales with respect to the underlying mesh size $\Delta_r$ as $\epsilon_r \sim \Delta_r^\gamma$. Thus, $\Delta_r \sim \epsilon_r^{\frac{1}{\gamma}}$ and $V_r \sim \epsilon_r^{-\frac{3}{\gamma}}$. As an implicit iterative method is used to compute the steady state flow field, the number of iterations or  \emph{pseudo time-steps} is given by $T_r$, which is dictated by the ability of the underlying linear solver to reach suitable convergence.

 On the other hand, the computational cost scales as a product of the number of cells and the number of iterations resulting in,  
\begin{equation}
\label{eq:ctr10}
C_r \sim f_r V_r T_r \sim f_r T_r \epsilon_r^{\frac{-3}{\gamma}} \ .
\end{equation}

Here, $C_r$ is the computational cost of generating a single realization (sample) of the flow field with the simulator $S_r$, given a fixed set of continuum inputs $v$ in Equation~\eqref{eq:x_cfd}, and $f_r$ denotes the \emph{flops per cell per step}, a key attribute of computational complexity that is discussed later in Appendix \ref{sec:appendixB}.

For an ideal steady-state RANS simulation, $T_r$ in \eqref{eq:ctr10} will be independent of mesh size as the goal of an ideal preconditioner is to keep the number of iterations constant, with respect to increasing mesh size. Such an ideal situation may be realized for Poisson-type solvers with (algebraic) multi-grid preconditioners. However, in practice, one has to use efficient ILU type preconditioners, where the number of iterations grows mildly with respect to increasing mesh size. To model these factors, we introduce an exponent $0 \leq \kappa << 1$ and set that $T_r \sim \Delta_r^{-\kappa}$. Empirically, an exponent of $\kappa \approx 0.2$ is observed for preconditioned Generalized minimal residual method (GMRES) solvers for CFD. With this ansatz, the asymptotic cost of data generation (per sample) scales as, 
\begin{equation}
\label{eq:ctr100}
C_r \sim f_r V_r T_r \sim f_r \epsilon_r^{\frac{-(3+\kappa)}{\gamma}} \ .
\end{equation}

\paragraph{Cost of model training} 
To train an AI model for acting as a surrogate for the RANS simulator $S_r$, we generate $N_r$ samples with the aforementioned data generation pipeline. The corresponding AI model of size $M_r$ (number of trainable parameters) is denoted by $S^\theta_r$. We call it the low-fidelity AI model as it is trained on low-fidelity RANS simulations. The cost of training this model, denoted by $C^T_r$ is given by the relation, 
 \begin{equation}
\label{eq:ctr2}
C^T_r \sim n f_e(I_r + M_r) N_r \ ,
\end{equation}
with $n$ being the number of strides (epochs) of the dataset during training, $f_e$ the flops per epoch (see below) and $I_r$ denoting the input size to the model. This input size is proportional to the the number of nodes or edges in (a downsampled version) of the underlying input graph. 
We are assuming a \emph{encode-process-decode} framework and an additive input size model for the encoder cost in the compute cost relation of Equation~\eqref{eq:ctr2}. Such an assumption is valid for UPT~\citep{alkin2024universal}, GAOT~\citep{wen2025geometry}, or GINO~\citep{li2023geometry} type models, i.e., models that use a locality based encoding that is similar to or based upon patching as introduced in ViTs~\citep{dosovitskiy2020image}). We see from Equation~\eqref{eq:ctr2} that the encoder costs scales as input size and the processor cost as model size. 

To further quantify the input size, we observe that the number of nodes are proportional to the number of spatial cells (volumes), given by $V_r \sim \Delta_r^{-3}$. The number of edges is also proportional to the same quantity. Therefore, the discussion above implies that $I_r \sim \epsilon_r^{-\frac{3}{\gamma}}$. It is essential to note that although they might be proportional, $I_r \neq V_r$, in general.

\paragraph{Scaling law for the low-fidelity AI model} 
Empirically, with some theoretical backing \cite{LMK1,deryck2022genericboundsapproximationerror}, one can show that there exist AI models such that the following scaling laws hold for approximating the low-fidelity RANS simulator, 
\begin{equation}
\label{eq:sl1}
\|S_{r} - S_r^\theta\| \sim \frac{1}{M_r^{\alpha}} + \frac{1}{N_r^{\beta}} \ .
\end{equation}
Here, $M_r$ is the model size and $N_r$ is the dataset size, measured here by the number of training samples. We are assuming an \emph{asymptotic regime} of large enough model and dataset sizes such that Equation~\eqref{eq:sl1} holds.

We can \emph{equilibrate} the two terms in Equation~\eqref{eq:sl1} by setting, 
\begin{equation}
\label{eq:eql1}
M_r \sim N_r^{\frac{\beta}{\alpha}} \ .
\end{equation}
Consequently, the resulting error is 
\begin{equation}
\label{eq:sl2}
\|S_{r} - S_r^\theta\| \sim  \frac{1}{N_r^{\beta}} \ .
\end{equation}

Now, we can obtain the following ground-truth error, 
\begin{equation}
\label{eq:sl3}
\begin{aligned}
\|S - S_r^\theta\| &\sim  \|S_{r} - S_r^\theta\|  + \|S_{r} - S_r^\theta\| \quad {\rm by~triangle~inequality} \\
&\sim \epsilon_r + \frac{1}{N_r^{\beta}}, \quad {\text{by Equation}}~\eqref{eq:rans}~{\text{and Equation}}~\eqref{eq:sl2} \ . 
\end{aligned}
\end{equation}
It is clear from the above equation that an AI model of size $M_r$ given by Equation~\eqref{eq:eql1} with $N_r$ training samples can obtain a test error of size $\epsilon_r$ (and no smaller error), provided that the dataset size and model size are chosen as, 
\begin{equation}
\label{eq:slr}
N_r \sim \epsilon_r^{-\frac{1}{\beta}}, \quad M_r \sim \epsilon_r^{-\frac{1}{\alpha}} \ .
\end{equation}

Substituting $M_r,N_r$ from Equation~\eqref{eq:slr} into Equation~\eqref{eq:ctr2} and also the asymptotic form of $I_r$, we obtain the model training cost as, 
\begin{equation}
\label{eq:ctr20}
C^T_r \sim n \epsilon_r^{-\frac{1}{\beta}} \left(\epsilon_r^{-\frac{1}{\alpha}} + \epsilon_r^{-\frac{3}{\gamma}} \right) \sim n f_e N_r\left(N_r^{\frac{\beta}{\alpha}} + N_r^{\frac{3\beta}{\gamma}} \right).
\end{equation}
Here, the last estimate follows from \eqref{eq:slr} by expressing compute in terms of the number of training samples, rather than the error. 

On the other hand, we observe from \eqref{eq:ctr100} that the data generation cost scales as, 
\begin{equation}
\label{eq:ctr3}
C^D_r \sim C_r N_r \sim f_r \epsilon_r^{-\left(\frac{(3+\kappa)} {\gamma}+\frac{1}{\beta}\right)} \sim f_r N_r^{1+\frac{(3+\kappa)\beta}{\gamma}} .
\end{equation}

These scaling laws are also summarized in Table \ref{tab:smscl1}
\begin{table}[htbp]
\caption{Complexity for training different AI models for CFD, comparing surrogates of low-fidelity (LF), high-fidelity (HF), and high-fidelity transient (HF-T) CFD simulators. All the quantities are defined in Section \ref{sec:smscl}.}
\centering
\setlength{\tabcolsep}{3pt} 
\begin{tabular}{|c|c|c|c|c|}
\hline
\textbf{Model} & \textbf{Error} & \textbf{Model size} & \textbf{Dataset size} & \textbf{Total compute} \\
\hline
$S^\theta_r$ (LF) & $\epsilon_r \sim N_r^{-\beta}$ & $ M_r \sim N_r^{\frac{\beta}{\alpha}}$ & $N_r$ &  $n f_e \left(N_r^{1+ \frac{\beta}{\alpha}} + N_r^{1+ \frac{3\beta}{\gamma}} \right) + f_r N_r^{1+\frac{(3+\kappa)\beta}{\gamma}} $ \\
\hline
$S^\theta_\ell$ (HF) & $\epsilon_\ell \sim N_\ell^{-\beta}$ & $ M_\ell \sim N_\ell^{\frac{\beta}{\alpha}}$ & $N_\ell$ &  $n f_e \left(N_\ell^{1+ \frac{\beta}{\alpha}} + N_\ell^{1+ \frac{3\beta}{\gamma}} \right) + f_\ell N_\ell^{1+\frac{4\beta}{\gamma}} $
\\
\hline
$S^\theta_t$ (HF-T) & $\epsilon_\ell \sim (\delta_K \tilde{N}_\ell)^{-\beta}$ & $ M_t \sim \tilde{N}_\ell^{\frac{\beta}{\alpha}}$ & $\tilde{N}_\ell$ &  $n f_e K\left(\tilde{N}_\ell^{1+ \frac{\beta}{\alpha}} + \tilde{N}_\ell^{1+ \frac{3\beta}{\gamma}} \right) + f_\ell \tilde{N}_\ell^{1+\frac{4\beta}{\gamma}} $
\\
\hline
\end{tabular}
\label{tab:smscl1}
\end{table}

\subsection{High-fidelity LES simulator} As a second configuration, we consider a high-fidelity LES simulator, that we denote by $S_\ell$ and assume that \begin{equation}
\label{eq:les}
\|S - S_\ell \| \sim \epsilon_\ell \ .
\end{equation}
Note that the output of $S_\ell$ is the time-averaged flow field $w_\ell$. The key difference between the RANS simulator $S_r$ and the LES Simulator $S_\ell$ lies in the fact that $\epsilon_\ell < \epsilon_r$ as, in general, LES models can capture flow phenomena more accurately than RANS models. Consequently, it also require a finer mesh size $\Delta_\ell$ for the underlying (mesh-converged) numerical method. 
An AI model $S_\ell^\theta$, of size $M_\ell$, can be trained on $N_\ell$ samples of this high-fidelity simulator. The corresponding scaling laws that relate error and cost can be derived, entirely analogously to the previous subsection, by simply replacing the index $r$  by the index $\ell$. 

Additionally, a key difference between the LES and RANS simulations lies in the time-stepping strategy. In contrast to the RANS configuration, where the steady state is directly computed using an iterative method, it is common to run a transient LES simulation and obtain the time-averaged or long-time limit fields as the steady state output. Consequently, we need to assume that $T_l$ $\sim \Delta_r^{-1}$, due to a CFL condition, resulting in Equation \ref{eq:ctr10a} (instead of \eqref{eq:ctr100} for RANS),

\begin{equation}
\label{eq:ctr10a}
C_\ell \sim f_\ell T_\ell V_\ell \sim f_\ell \epsilon_l^{\frac{-4}{\gamma}} \ .
\end{equation}

The final scaling laws for this model, expressed in terms of the dataset size, are summarized in Table \ref{tab:smscl1}.

\subsection{Implications of the scaling laws.}
Given the above scaling laws, we can already derive the following implications, 
\paragraph{Processor dominates encoder in training costs.} We see from Table \ref{tab:smscl1} that the processor cost scales as $N_{\ell,r}^{1+\frac{\beta}{\alpha}}$ and the encoder cost scales as $N_{\ell,r}^{1 + \frac{3\beta}{\gamma}}$. Hence, as long as $3\alpha < \gamma$, the processor dominates the encoder in terms of their relative contributions to the training cost. This regime clearly holds as $\gamma \approx 1$ (rate of convergence of numerical method) while $\alpha \in [0.2,0.3]$, as empirically observed in \cite{herde2024poseidon,wen2025geometry} and references therein.

\paragraph{Relative balance between model training costs and data generation costs.} We begin with an analysis of the relative contributions of the costs of model training and data generation in the case of the high-fidelity AI model. Under the realistic assumption that the processor dominates the encoder in the model training costs, we obtain the following balance relation from Table \ref{tab:smscl1}
\begin{equation}
\label{eq:rat1}
\frac{C^T_{\ell}}{C^D_\ell} \sim  \frac{nf_e}{f_\ell} N_\ell^{\beta \left(\frac{1}{\alpha} - \frac{4}{\gamma}\right)}\ .
\end{equation}
Hence, we have the following two cases, 
\begin{itemize}
\item [(i.)] $4\alpha \geq \gamma$: In this case, the error of the AI model converges quite fast in model size. Consequently, it follows from Equation~\eqref{eq:rat1} that $C^T_\ell < C^D_\ell$ as long as $nf_e < f_\ell$. Hence, data generation costs will dominate model training costs in this case.  
\item [(ii.)] $4\alpha < \gamma$: In this case, the error of the AI model converges relatively slowly in model size. Consequently, $C^T_\ell < C_\ell^D$ only if,
\begin{equation}
\label{eq:rat2}
N_\ell \leq \left( \frac{f_r}{n f_e}\right)^{\frac{\alpha \gamma}{\beta(\gamma - 4\alpha)}} \ .
\end{equation}
Beyond this critical number of training samples, the model generation costs will dominate as long as $nf_e \leq f_r$.
\end{itemize}

Thus, a careful analysis of the scaling laws derived here show that in the realistic case of slow convergence with respect to model size, we can expect that the data generation costs will dominate the model training costs till a critical number of samples are reached. Once this limit is breached, the model training costs will start to dominate the cost of data generation. However, the exact number of training samples is clearly problem dependent. 

A similar analysis can be repeated for the low-fidelity model, leading to the same conclusions but the case separation condition reads $(3+\kappa)\alpha < \gamma$ in this case. This condition is even more likely to hold as $\kappa << 1$. 

\paragraph{Training with high-fidelity simulations is more accurate but more expensive than low-fidelity simulations.} It is clear from Table \ref{tab:smscl1} that one can at most reach a test error of size $\epsilon_r$ with the AI model $S^\theta_r$, trained on the low-fidelity simulations. One cannot reach a test error of size $\epsilon_\ell$ with this model. 

On the other hand, one can reach a test error of size $\epsilon_\ell$ with the AI model $S^\theta_\ell$, trained on high-fidelity simulations. However, this model will clearly be more expensive. 

To compare the costs, we assume, for the simplicity of analysis, that the data generation costs dominate over the model training cost and divide the corresponding entries of the last columns in Table \ref{tab:smscl1}, using the simplified cost model of Equation~\eqref{eq:ctr10} to obtain
\begin{equation}
\label{eq:rat3}
\frac{C_\ell}{C_r} \sim \left(\frac{\epsilon_r}{\epsilon_\ell}\right)^{\left(\frac{1}{\beta} + \frac{3+\kappa}{\gamma}\right)} + \left(\frac{1}{\epsilon_\ell}\right)^{\left(\frac{1}{\beta} + \frac{1-\kappa}{\gamma}\right)} \ .
\end{equation}
Thus, for any values of $\beta,\gamma$, $0 \leq \kappa \leq 1$, $C_\ell > C_r$ as $1 > \epsilon_r > \epsilon_\ell$. In fact, the cost differential between training the two models grows as a (large) power of the relative error ratio. A similar analysis can be repeated in the case where the model training costs dominate. 
\subsection{High-fidelity transient LES simulations}
We recall that the high-fidelity LES simulation actually approximates a transient flow field $\{\vec{u}(t_n)\}$, sampled at time snapshots $t_n$ with $n=1,2,\ldots,N$. Thus, each high-fidelity simulation actually contains the \emph{trajectory} of the evolution of the fluid state. These simulations are either averaged in time or a long-time limit is taken to obtain the steady-state output, as considered in the previous section. This observation naturally raises the question: {\it can one leverage the fact that the high-fidelity simulations provide transient information to scale the high-fidelity AI model better?} We attempt to answer this question below. 

First, we observe that the underlying learning problem is \emph{different} for both cases. The underlying operator in the low fidelity case $S: v \mapsto w$, provides a single functional output $w$ for the given input $v$. 
On the other hand, the underlying operator in the transient case is given by $S_t: v \mapsto w(t)$, which maps the input $v$ (which includes the initial conditions $v(0)$) into a time-dependent output $w(t)$, for $t \in [0,T]$. In practice, this output is sampled at discrete time points $t_n$, implying that the operator maps $S_t: v \mapsto \{w(t_n)\}$, for $n =1,2,\ldots,N$. By concatenating all the snapshots into a vector $\vec{w} = \{w(t_n)\}$, we identify the underlying operator as $S_t: v \mapsto \vec{w}$. Hence, there is a clear distinction between learning $S$ and learning $S_t$. They are related by 
\begin{equation}
\label{eq:tr1}
\lim\limits_{t \rightarrow \infty} S_t = S, \quad {\rm or} \quad \frac{1}{T} \int\limits_0^T S_t dt \sim S \ .
\end{equation}

We assume that the high-fidelity LES simulations, denoted by $S^\ell_t$ can approximate $S_t$ in terms of 
\begin{equation}
\label{eq:tr2}
\| S_t - S^\ell_t \| \sim \epsilon_\ell \ .
\end{equation}
Here, the norm $\|.\|$ clearly involves time-dependent quantities and requires integrating over time. The corresponding data to learn $S_t$ is given by $(v_i,\vec{w}_i)$ for all $1 \leq i \leq N$, with $\vec{w}_i = \{S^\ell_{t_n}(v_i)\}_{n=1}^N$. We observe that the trajectory data need not be saved at every single time-step of the underlying numerical simulation, but at representative time steps. 

The AI model trained on this time-dependent trajectory data is denoted as $S_t^\theta$. During its training, it sees the trajectories either in an autoregressive manner \citep{bi_accurate_2023,lam2023graphcast_new,brandstetter2022message,bodnar_aurora_2024} or using specialized techniques such as \emph{all2all} training \citep{herde2024poseidon}, in which one can train the model to minimize the \emph{loss function}
\begin{equation}
\label{eq:tr3}
\sum\limits_i \sum\limits_{j < k} {\mathcal L}\left(w^i(t_k), S^\theta_t(w^i(t_j),t_k-t_j) \right) \ .
\end{equation}
Thus, given the data at time step $t_j$ and the lead time $t_k - t_j$, the AI model is trained to predict the ``velocity field'' $w^i(t_k) = S^\ell_{t_k}(u^i)$, at a later time $t_k$. In principle, this implies that for the same sample, indexed here by $i$, the AI model $S^\theta_t$ sees $K$ effective copies of the data with  $1 \leq K \leq (\tilde{N}_{\ell} (\tilde{N}_\ell + 1))/2$

We can hypothesize the following (corrected) scaling law for sufficiently large model size (see also \cite{herde2024poseidon}): 
\begin{equation}
\label{eq:tr4}
\|S_{t}^\ell - S_t^\theta\| \sim \frac{1}{\delta_K^{\beta} \tilde{N}_\ell^{\beta}} \ .
\end{equation}
with $\delta_K$ being a compression term that measures how fewer samples are needed to obtain the same error when the model has been trained on full trajectories, rather than only on the steady state or time-average.  

Thus the number of effective samples and model size to obtain the desired error scale as,
\begin{equation}
\label{eq:tr5}
\tilde{N}_\ell \sim \frac{\epsilon_\ell^{\frac{-1}{\beta}}}{\delta_K}, \quad M_t \sim \tilde{N}^{\frac{\beta}{\alpha}}.
\end{equation}

The cost of training this transient model $S^\theta_t$ is given by
\begin{equation}
\label{eq:tr6}
\tilde{C}_T \sim n f_e(M_t+I_\ell) (K\tilde{N}_\ell) \sim n \frac{K}{\delta_K} \left(\epsilon_\ell^{-\left(\frac{1}{\alpha} + \frac{1}{\beta}\right)} + (\epsilon_\ell^{-\left(\frac{3}{\gamma} + \frac{1}{\beta}\right)}\right) \sim n f_e K\left(\tilde{N}_\ell^{1+ \frac{\beta}{\alpha}} + \tilde{N}_\ell^{1+ \frac{3\beta}{\gamma}} \right)  \ .
\end{equation}
On the other hand, the data generation cost scales as,
\begin{equation}
\label{eq:tr7}
\tilde{C}_D  \sim  f_\ell \tilde{N}_\ell^{1+\frac{4\beta}{\gamma}} \ .
\end{equation}
The total cost is summarized in the third-column of Table \ref{tab:smscl1}.

We readily observe from Table \ref{tab:smscl1} that the high-fidelity transient model is clearly cheaper than the high-fidelity time-averaged model, when the data generation costs dominate as $\tilde{N}_\ell < N_\ell$. In the asymptotic regime where the model training costs, we see that the high-fidelity transient model continues to be cheaper to train than the high-fidelity time-averaged model as long as $\delta_K^{1 + \frac{\beta}{\alpha}} > K$. 

\section{Constraining the theory with concrete numbers}
The theory presented thus far yields interesting qualitative conclusions regarding the training of a foundation AI model for CFD; however, we also aim to establish quantitative cost estimates. Obtaining precise figures is challenging, as the bounds provided in Table \ref{tab:smscl1} hold only up to a generic constant (denoted by $\sim$). Despite this limitation, we offer representative values based on reasonable assumptions detailed in Appendices \ref{sec:appendixB} and \ref{sec:appendixC}. While these calculations may deviate from exact costs, they provide order-of-magnitude estimates intended to guide future research.

\subsection{Explicit compute for representative configurations.}

We start by focusing on the high-fidelity LES simulation which yields a time-averaged field as its learning task (middle row of Table \ref{tab:smscl1}). Using a representative simulation of a common low-speed subsonic external aerodynamics simulation of a car or an airplane \citep{cetin2023b,ashton2023b}, we require $V_\ell = 5 \times 10^8$ spatial cells and $T_\ell = 2 \times 10^5$ time-steps or iterations to achieve the desired accuracy (see Table \ref{tab:smscl2}) \footnote{We accept that these numbers are use-case dependent, with some simulations requiring larger computational resources and others requiring less.}.We assume the use of an explicit cartesian WMLES calculation that is representative of the state-of-the-art simulations that typically improves correlation to ground truth experimental data \citep{cetin2023b,ashton2023b} compared to RANS. See Appendix \ref{sec:appendixB} for further details on the derivation of this flops per cell per step number that is based upon an explicit cartesian solver. Entirely for demonstrative purposes, we assign a relative error level of $\epsilon_\ell \sim 3\%$, for the time-averaged (and transient) flow fields. This number is consistent with the requirement that the relative error for the integral quantities (drag, lift etc) is approximately $1\%$.  

Similarly, for the low-fidelity RANS simulation for the same use case, we consider a mesh of $V_r = 10^8$ points and the number of iterations to reach steady state is given by $T_r = 2000$. We also assume the use of a unstructured implicit steady-state CFD solver using a GMRES approach using a RANS model such as the $k-\omega$ model \citep{Menter1994}. 
Thus, the effective mesh sizes of the low-fidelity and high-fidelity simulations are given by $\Delta_r \approx 1.71 \Delta_\ell$, noting though that the high-fidelity is solving a time-accurate solution. 

Assuming a widely observed first-order of mesh convergence of numerical methods by setting $\gamma \approx 1$, we see that the the field error is $\epsilon_r \approx 5\%$ \footnote{We note that for industrial use-cases the error level between different CFD inputs choices (WMLES vs RANS) will be case dependent.}.

\paragraph{Refined estimates of training cost.} From \eqref{eq:ctr2}, we have assumed how an additive model for how inputs are encoded and then processed in an AI model, translates into compute. However, to relate to practical use cases and provide reasonable estimates of the model training cost, we need to refine this formula further. To this end, we assume the following relation, 
\begin{equation}
\label{eq:mf1}
C^T \approx nNf_e(IC_E + \tau M),  
\end{equation}
with model training cost $C^T$, $N$ being number of samples, $n$ being the number of epochs, $I$ input size and $M$ model size. In addition, we also need to introduce the number of embedding channels $C_E$ in the encoder and decoder and the number of \emph{latent tokens} $\tau$ that is the input to a processor. This refined estimate holds for model architectures such as UPT, GINO and GAOT and is consistent with the asympototic estimates of Section E.1 and E.2 of \cite{wen2025geometry}. The new factor introduced in Equation~\eqref{eq:mf1} is $f_e$, which we term as \emph{flops per training epoch}, that allows us to estimate compute (in flops) more accurately. 

For the transient case, we can readily adapt Equation~\eqref{eq:mf1} to, 
\begin{equation}
\label{eq:mf1}
C^T \approx n\tilde{N}f_eK(IC_E + \tau M) \ ,  
\end{equation}
to account for the fact that $K$ input-output pairs are seen by the model (per epoch) during training.

\paragraph{Input size.} Once the high-fidelity LES fields have been generated at a resolution of $500\hspace{1pt}\text{M}$, we can readily \emph{downsample} the fields to a coarser mesh, such that the resulting downsampling or interpolation error is smaller than $3\%$. Again, for demonstrative purposes, we consider a compression factor of $64$ (corresponding to a factor of $4$ compression in each direction). We note that in practice this downsampling would be done in a non-uniform manner to account for areas that require more resolution e.g boundary layers. Moreover, the downsampled nodes and edges are randomly chosen per training epoch and can be changed from epoch to epoch, ensure sufficient spatial coverage of the whole domain during the training process. The resulting mesh has $7.81\hspace{1pt}\text{M}$ input points or nodes and assuming that the number of edges scales as $6$ times the number of nodes (for a cartesian mesh), the number of edges is now $46.86\hspace{1pt}\text{M}$. Now the edge features have to be transformed by an encoder MLP, which we assume is $C_E = 32^2 = 1024$ channels. 

We can apply similar considerations to compute the input size of the low-fidelity model by applying a downsampling ration of $64$, leading to an input graph with $I= 9.25\hspace{1pt}\text{M}$ edges and an encoder MLP of size $C_E = 32^2 = 1024$ channels. 

\paragraph{Dataset size.} To compute the number of samples in each setting, we need an estimate of the scaling exponent $\beta$, by which the test error scales with respect to the number of samples. Statistical learning theory suggests an upper bound of $\beta < 0.5$ \cite{LMK1,deryck2022genericboundsapproximationerror} and a slightly smaller rate has been observed in practice \cite{herde2024poseidon,wen2025geometry,rigno}. To provide a representative rate, we will use the scaling exponent for the Poseidon foundation model~\citep{herde2024poseidon}, where from Figure 23 (bottom right), this rate is given by $\beta \approx 0.43$. 

This exponent does not suffice to calculate an estimate of the number of samples needed to attain the error level of approximately $3\%$ for the high-fidelity model as the proportionality constants are missing. Again, we can estimate them from Figure 23 of the Poseidon foundation model, where $80\hspace{1pt}\text{K}$ samples where necessary to obtain a test error of $9\%$. Extrapolating from this error level with an exponent of $0.43$, leads to $1.02\hspace{1pt}\text{M}$ training samples, which we approximate by $1\hspace{1pt}\text{M}$ samples.  

For the low-fidelity case, we know that the desired error level is $5\%$. Again, with $\beta = 0.43$, and the fact that an error of $3\%$ required approximately $1\hspace{1pt}\text{M}$ samples, we extrapolate to obtain a figure of $325\hspace{1pt}\text{k}$ samples (approximately) in this case. 

For the high-fidelity transient case, we also need to estimate the compression ration $\delta_K$ and the effective number of time snapshots $K$ that are used for training the high-fidelity transient model. We assume that we save $100$ time steps per high-fidelity transient sample. With the \emph{all2all} training strategy of \cite{herde2024poseidon}, we can use at most $5050$ input-output pairs per sample. However, such an extreme use of transient information is excessive. Instead, to keep computational costs reasonable and ensure generalization, we will use approximately the configuration of \cite{herde2024poseidon} by using $50$ (randomly selected) data pairs per sample per epoch, leading to setting $K=50$. We note that a different set of $50$ input-output pairs per sample can be sampled for a different epoch, ensuring proper coverage of all the transient information in the dataset. For the compression ratio $\delta_K$, we follow the observations of \cite{herde2024poseidon} Figure 46 by setting $\delta_K = 5$. Thus the number of training examples in the transient high-fidelity case is given by $\tilde{N}_\ell = 200\hspace{1pt}\text{k}$, as reported in Table \ref{tab:smscl2}.

In addition we take a second, independent approach to reach an estimate on the number of samples required. As shown in Appendix \ref{sec:appendixC}, we consider a broad range of possible geometries and flow conditions across the entire spectrum of CFD. We estimate in the order of $2.25M$ samples.  

\paragraph{Model size.} To compute the model size, we rely on the relation Equation~\eqref{eq:eql1}, which relates model size to the number of training samples. To this end, we need to estimate the scaling exponent $\alpha$. There are very few estimates of $\alpha$ available in the literature but we can use the value of $\alpha$ from \cite{herde2024poseidon} (Figure 22), as observed in the case of the Poseidon PDE foundation model. This value of $\alpha \approx 0.24$ is also consistent with the observations of \cite{wen2025geometry}. Using this value of $\alpha$ and $\beta = 0.43$ and the number of samples $N_\ell=10^6$, we arrive at a model size of $\approx 56\hspace{1pt}\text{B}$ parameters, which is reported in Table \ref{tab:smscl2}. Repeating the same calculation steps for the low-fidelity case leads to a model size of approximately $7\hspace{1pt}\text{B}$ parameters. 

To estimate $\tau$, the number of \emph{latent tokens} that are fed into the processor of our AI model, we use the ball-park estimates of \cite{wen2025geometry} where the latent grid had size of $64^3$ and the patch size for a ViT backbone was $p=2$, resulting in a total of $\tau = 32768$ tokens per sample. This also coincides with Figure 7 of~\citet{alkin2025ab}, where it is shown that on the DrivAerML dataset models start to overfit when using in the order of $100\hspace{1pt}\text{k}$ anchor tokens.
We keep the same number of latent tokens for the low-fidelity model too. 

\paragraph{Estimates of Flops per epoch.} To compute estimates of flops per training epoch $f_e$, we base our calculations on a training run with a GAOT type transformer model \citep{wen2025geometry}, which is trained using a GH200 NVIDIA GPU with BF16 precision. In this run, a model size of $1.4\hspace{1pt}\text{B}$ parameters was used with an input size of $3\hspace{1pt}\text{M}$ edges and embedding channel dimension of $C_E = 1024$. The underlying structured latent grid had $64^3$ tokens. Two configurations were tested: one with a standard ViT~\citep{dosovitskiy2020image} with a patch size of $4$, leading to $\tau = 4096$ tokens and another with a highly-optimized Swin transformer~\citep{liu2021swin} processor, which supported a patch size of $2$. Based on a (average) sustained performance of $1.19 \times 10^{18}$ flops and theoretical estimates of $2.87 \times 10^{16}$ flops and $2.3 \times 10^{17}$ flops for the two configurations, we arrive at $f_e \approx 5.2$ for a patch size of $p=2$ and $f_e \approx 41.5$ for a patch size of $p=4$, respectively.  

\paragraph{Number of epochs.}
Contrary to LLMs where training for one epoch corresponds to seeing all training data, CFD data can be heavily subsampled without losing information~\citep{alkin2025ab}. However, several combinations of subsampled inputs are needed to reflect the full physics of the data. Thus, several epochs are needed to exploit the full information contained in the simulations.

\begin{table}[htbp]
\centering
\begin{threeparttable}
\caption{Concrete numbers for developing a CFD foundation model.}
\label{tab:smscl2}
\setlength{\tabcolsep}{4pt} 

\begin{tabular}{@{}l c c c@{}}
\toprule
& \textbf{Low-fidelity} & \textbf{High-fidelity} & \textbf{Transient high-fidelity} \\ 
\midrule
\multicolumn{4}{@{}l}{\textit{Data generation}} \\
Training samples ($N$)          & $3.25 \times 10^5$  & $10^6$               & $2 \times 10^5$ \\
Number of cells ($V$)           & $10^8$              & $5\times 10^8$       & $5 \times 10^8$ \\
Number of steps ($T$)           & $2000$              & $200000$             & $200000$ \\
Flops per cell per step ($f$)       & $40000$             & $2800$               & $2800$ \\
Compute flops ($C$)         & $2.6 \times 10^{21}$& $2.8 \times 10^{23}$ & $5.6 \times 10^{22}$ \\
GPU hours ($C_{h}$) \tnote{a}  & $1.81 \times 10^{5}$& $6.48 \times 10^{6}$ & $1.3 \times 10^{6}$ \\
Compute cost ($C_{c}$) \tnote{b} & \$$1.4\hspace{1pt}\text{M}$            & \$$51.8\hspace{1pt}\text{M}$              & \$$10.3\hspace{1pt}\text{M}$ \\
Storage ($S$) \tnote{c}       & $650\hspace{1pt}\text{TB}$               & $2\hspace{1pt}\text{PB}$                  & $400\hspace{1pt}\text{TB}$ \\
Yearly storage cost ($S_c$) \tnote{d} & \$$78\hspace{1pt}\text{k}$         & \$$240\hspace{1pt}\text{k}$               & \$$48\hspace{1pt}\text{k}$ \\
\midrule
\multicolumn{4}{@{}l}{\textit{Model training}} \\
Data scaling exponent ($\beta$) & $0.43$              & $0.43$               & $0.43$ \\
Effective copies ($K$)          & --                  & --                   & $50$ \\
Compression ratio ($\delta_K$)  & --                  & --                   & $5$ \\
Training samples ($N$)          & $3.25 \times 10^5$  & $10^6$               & $2 \times 10^5$ \\
Model scaling exponent ($\alpha$) & $0.24$            & $0.24$               & $0.24$ \\
Model size ($M$)                & $7.5 \times 10^9$   & $5.62\times 10^{10}$ & $3.15\times 10^9$ \\
Epochs ($n$)                    & $100$               & $100$                & $100$ \\
Flops ($f_{e}$)          & $5.2$              & $5.2$              & $5.2$ \\
Input size ($I$)                & $9.3 \times 10^6$   & $4.7 \times 10^{7}$ & $4.7 \times 10^{7}$ \\
Encoding Channel Dimension ($C_E$)& $1024$  & $1024$  & $1024$ \\
Patch Size ($p$)          & $2$              & $2$              & $2$ \\
Number of Latent Tokens ($\tau$) & $32768$ & $32768$  & $32768$ \\
Training flops ($C^T$)          & $4.16\times 10^{22}$& $9.58 \times 10^{23}$& $5.36 \times 10^{23}$ \\
GPU hours ($C^{Th}$) \tnote{e}  & $3849$                 & $8.8 \times 10^{4}$ & $4.96 \times 10^{4}$ \\
Compute cost ($C^{Tc}$) \tnote{b} & \$$30.7\hspace{1pt}\text{k}$            & \$$709.7\hspace{1pt}\text{k}$                & \$$397.2\hspace{1pt}\text{k}$ \\
\bottomrule
\end{tabular}

\begin{tablenotes} \footnotesize
\item [a] Assuming NVIDIA GB200 in FP32 ($80$ TFLOPS) with $15$\% (explicit WMLES) and $5$\% (implicit RANS) sustained flops usage.
\item [b] Assuming \$$8$ cost per hour for a single NVIDIA GB200 GPU.
\item [c] Assuming $2\hspace{1pt}\text{GB}$ for FP32 with 10 variables stored on downsampled mesh.
\item [d] Assuming \$$0.01$ per GB per month.
\item [e] Assuming NVIDIA GB200 in BF16 ($5$ PFLOPS) with $60$\% sustained flops usage.
\end{tablenotes}
\end{threeparttable}
\end{table} 

The results of the afore described explicit configurations are summarized in Table \ref{tab:smscl2}, where the report the total compute (in terms of data generation as well as model training) for the three scenarios considered here, namely the low-fidelity steady state, high-fidelity time-averaged and the high-fidelity transient cases. We also provide a dollar cost for each of the models. This table provides a principled estimate of the costs for training CFD Foundation models in the regime described above. We observe from this table that all the implications, suggested by the theory in the previous section, seem to hold. In particular, the cost of the low-fidelity model is significantly lower than the high-fidelity ones. Also, data generation costs do outweigh model training costs in this regime of less than $17\hspace{1pt}\text{M}$ samples. 
\begin{figure}    
\centering
    \begin{subfigure}[t]{0.48\textwidth}
        \centering
        \includegraphics[width=\textwidth]{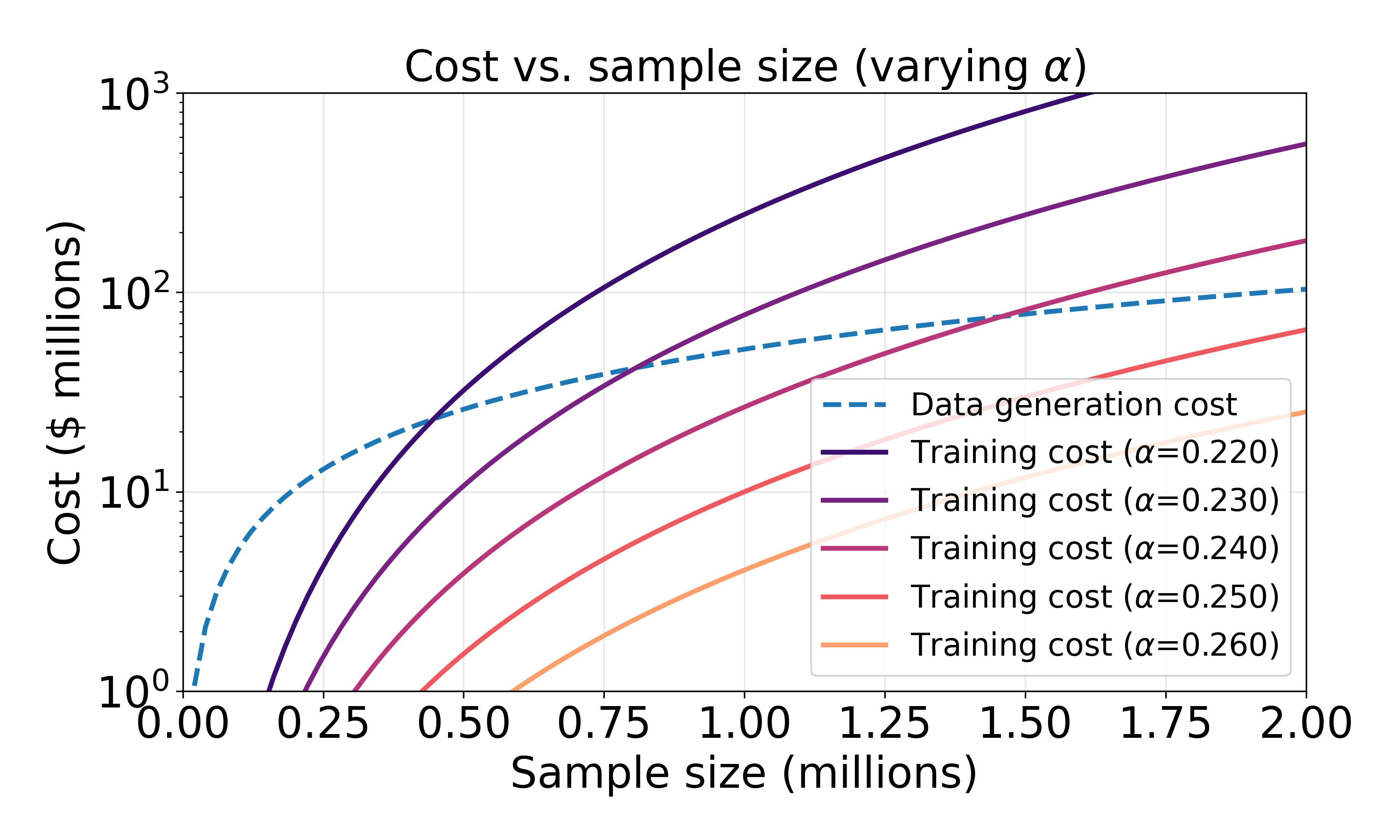}
        \caption{\textbf{Varying $\boldsymbol{\alpha}$}:  $\beta=0.425$, Epochs=100.}
        \label{fig:alpha}
    \end{subfigure}
    \hfill
    \begin{subfigure}[t]{0.48\textwidth}
        \centering
        \includegraphics[width=\textwidth]{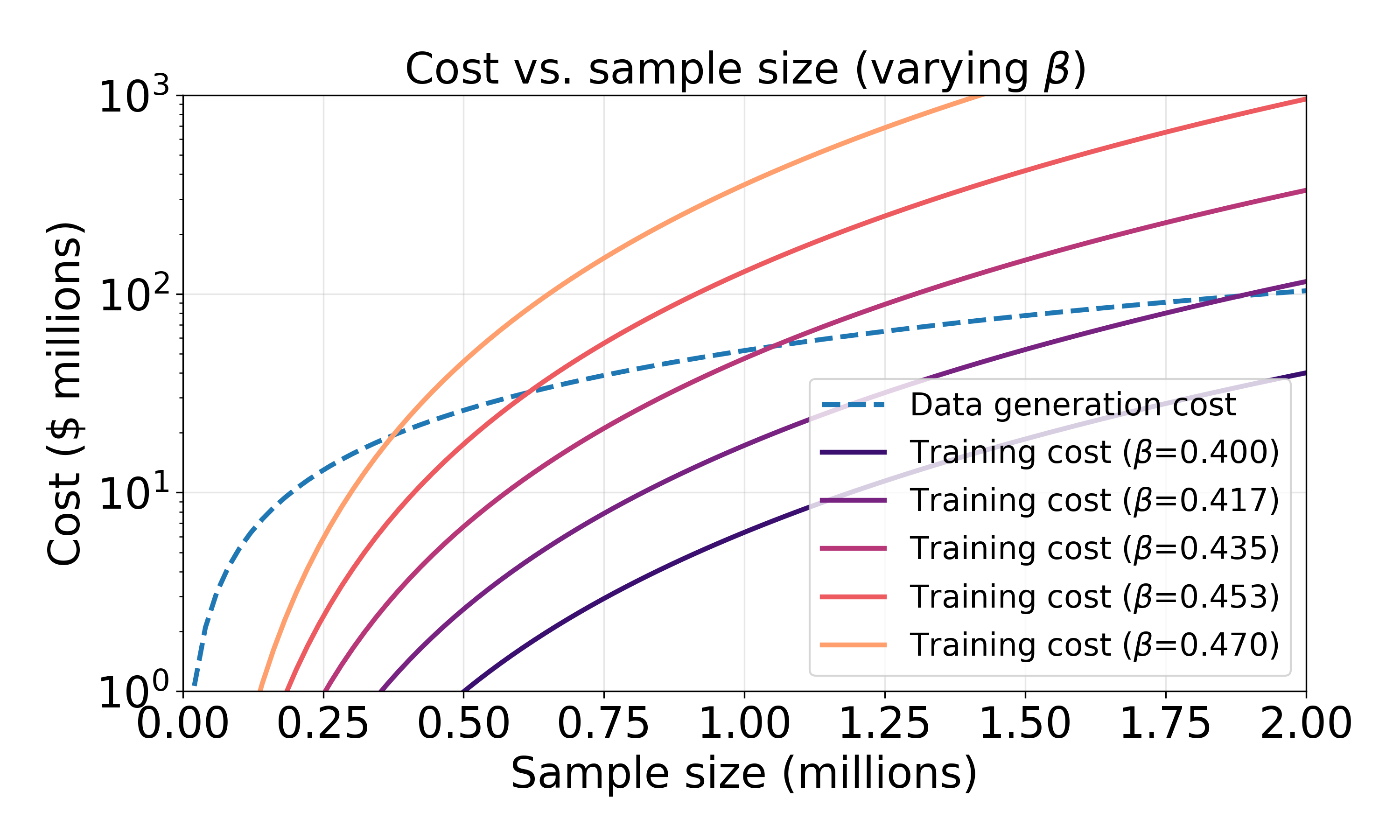}
        \caption{\textbf{Varying $\boldsymbol{\beta}$}:  $\alpha=0.24$, Epochs=100.}
        \label{fig:beta}
    \end{subfigure}
    
    \vspace{0.25cm} 
    
    \begin{subfigure}[t]{0.48\textwidth}
        \centering
        \includegraphics[width=\textwidth]{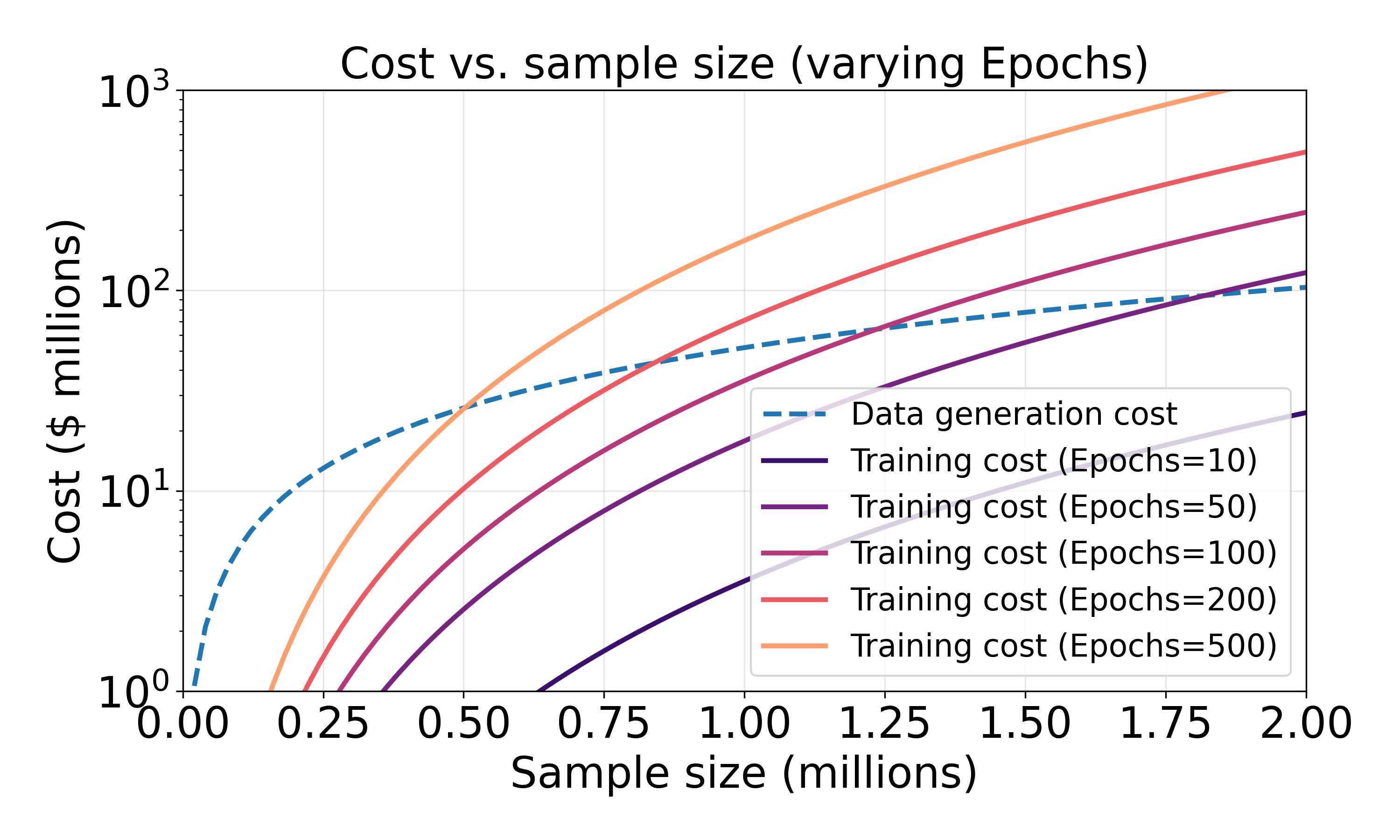}
        \caption{\textbf{Varying epochs}: $\alpha=0.24, \beta=0.43$.}
        \label{fig:epochs}
    \end{subfigure}
    \hfill
    \begin{subfigure}[t]{0.48\textwidth}
        \centering
        \includegraphics[width=\textwidth]{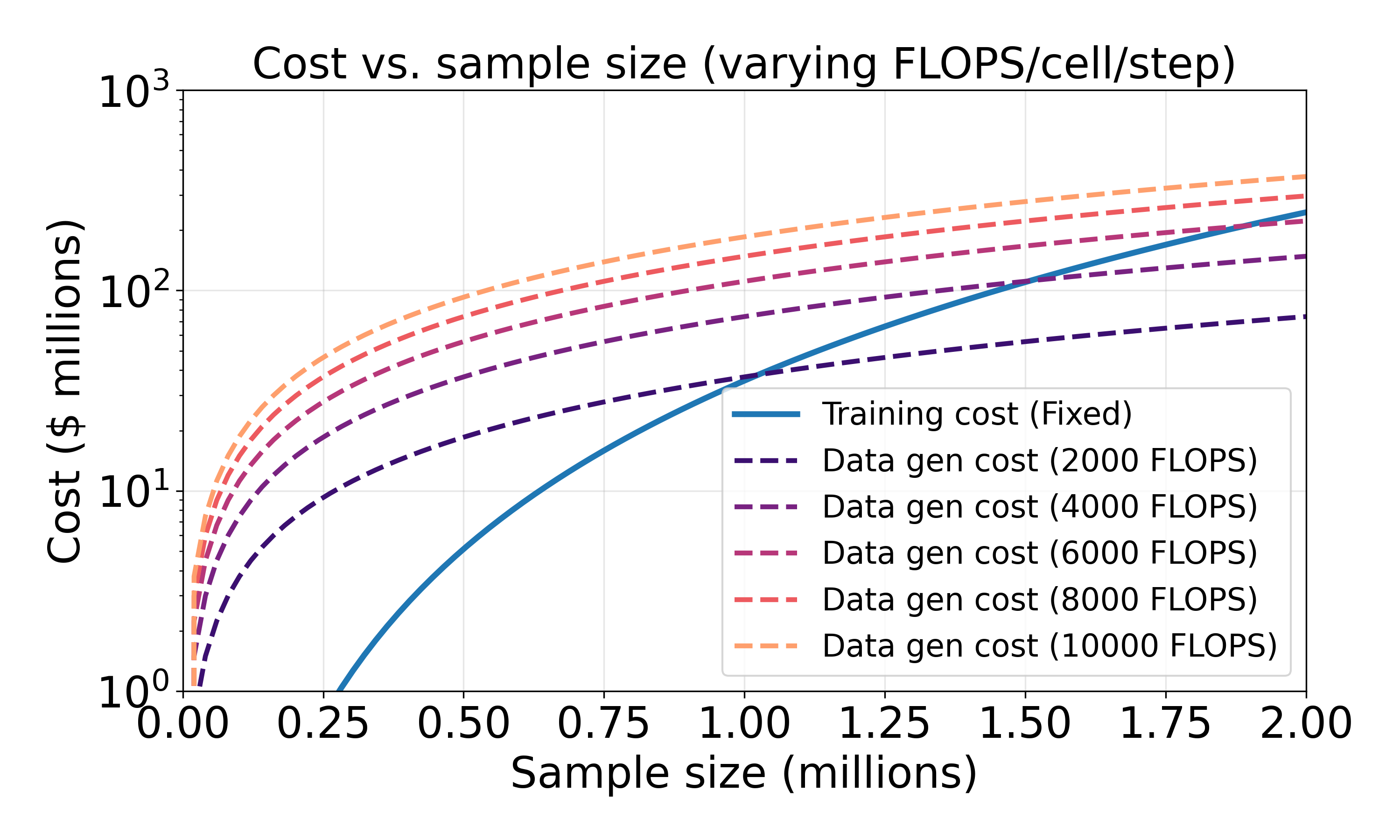}
        \caption{\textbf{Varying FLOPS per step per cell}: Fixed training with Epochs=100 $\alpha=0.24$ and $\beta=0.43$.}
        \label{fig:datagen}
    \end{subfigure}

    \vspace{0.25cm} 
    
    \begin{subfigure}[t]{0.48\textwidth}
        \centering
        \includegraphics[width=\textwidth]{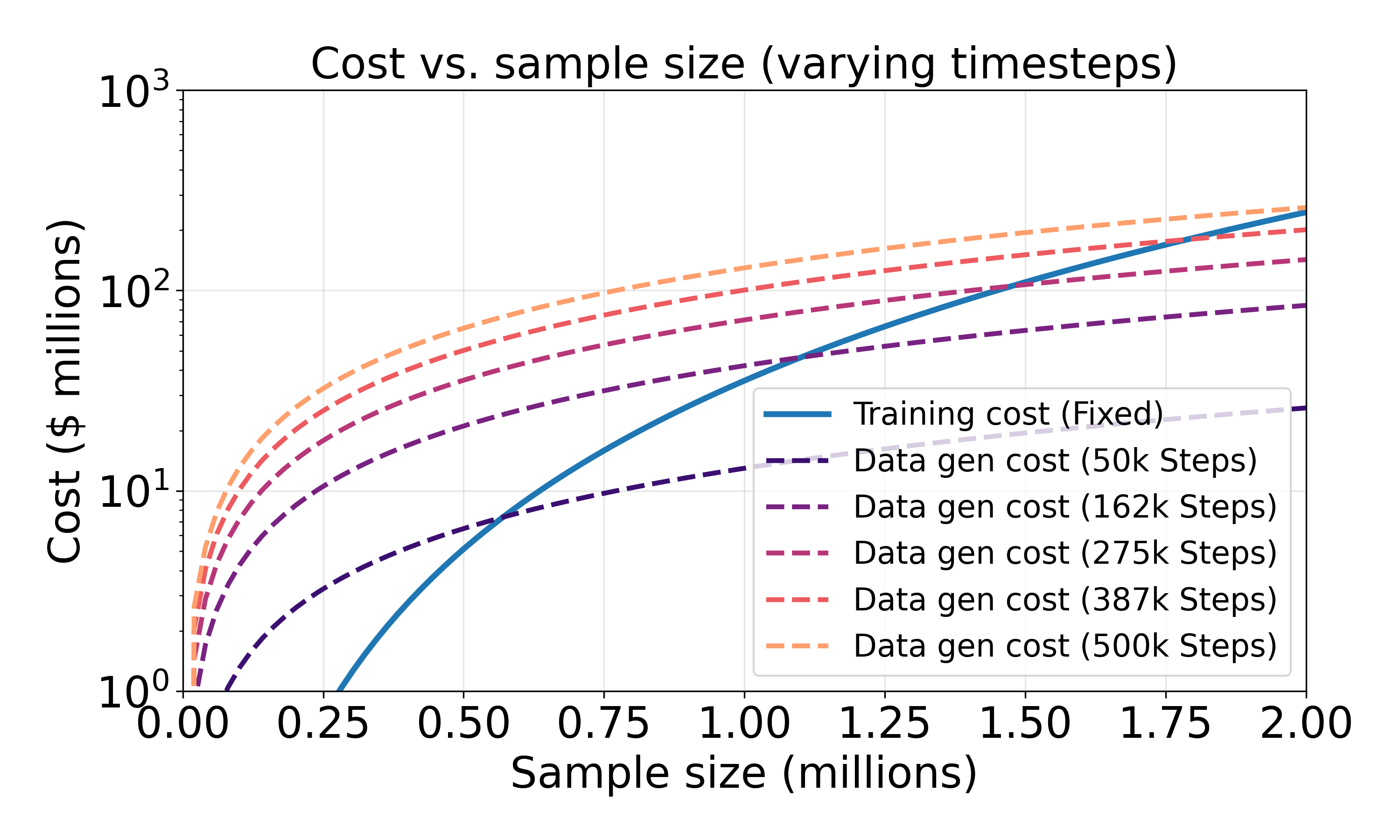}
        \caption{\textbf{Varying time steps}: Fixed training with Epochs=100 $\alpha=0.24$ and $\beta=0.43$.}
        \label{fig:timesteps}
    \end{subfigure}
    \hfill
    \begin{subfigure}[t]{0.48\textwidth}
        \centering
        \includegraphics[width=\textwidth]{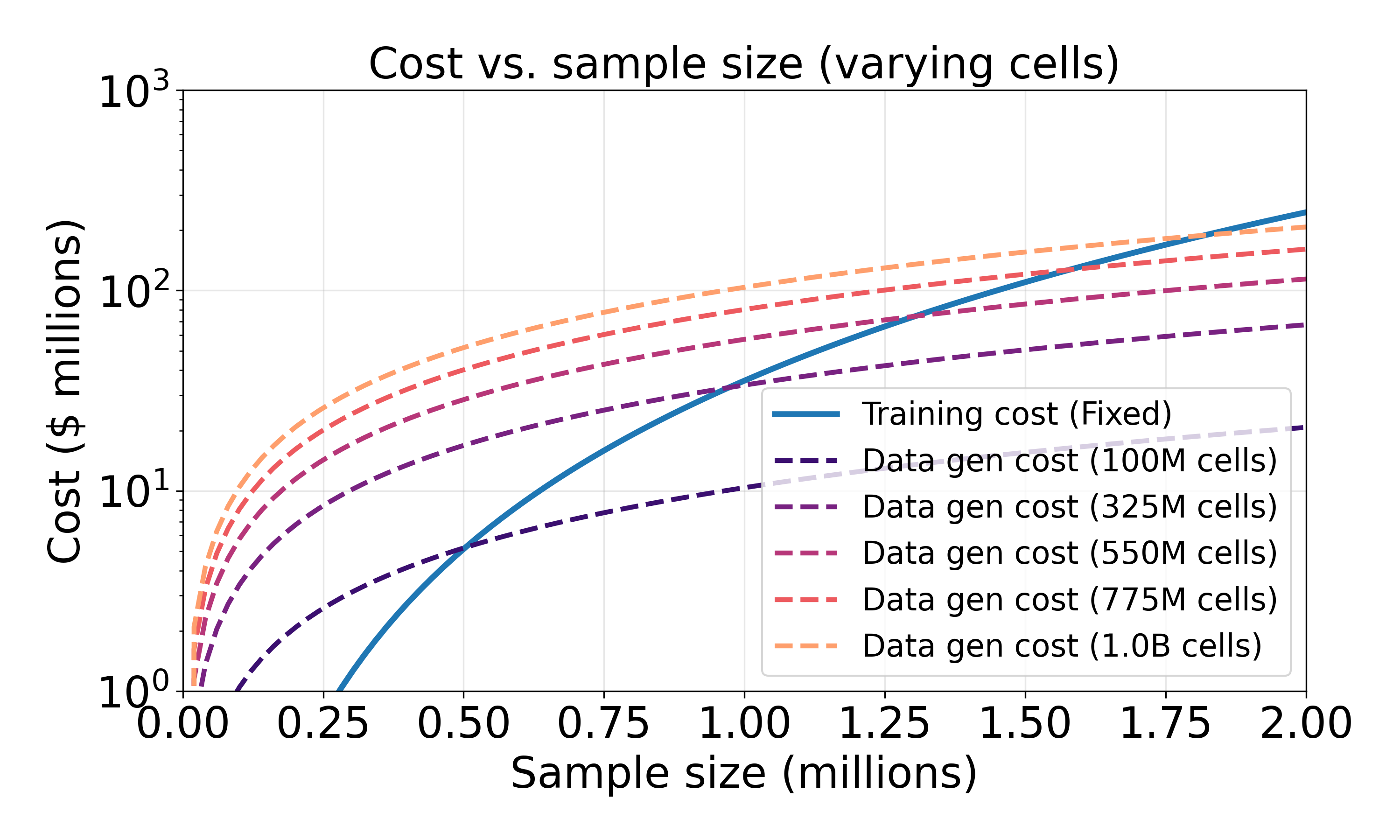}
        \caption{\textbf{Varying cell count}: Fixed training with Epochs=100 $\alpha=0.24$ and $\beta=0.43$.}
        \label{fig:cell}
    \end{subfigure}    
    
    \caption{Comparative analysis of data generation and model training costs ($y$-axis in \$ millions) versus sample size ($x$-axis in millions) for the transient high-fidelity case. Source code provided in Appendix \ref{sec:code}}
    \label{fig:grid}
\end{figure}

\subsection{Assessing scaling behavior}
To further explore the practical implications of our scaling theory, Figure \ref{fig:grid} visualizes the interplay between data generation ($C^D$) and model training ($C^T$) costs across varying sample sizes and parameter configurations for the transient high-fidelity case. As suggested by the theory, a consistent trend emerges across all parameter variations: while data generation represents the dominant computational burden for smaller sample regimes (typically $N < 10^6$), the superlinear scaling of training costs -- driven by the coupled growth of model size $M$ and dataset size $N$ required to satisfy Equation \ref{eq:slr} eventually overtakes generation costs. Subplots \ref{fig:alpha} and \ref{fig:beta} highlight the sensitivity of this crossover point to the scaling exponents; notably, a lower model scaling exponent $\alpha$ (indicating slower error convergence per parameter) significantly inflates the required model size, causing training costs to become the primary bottleneck at much lower sample counts. Conversely, variations in CFD input parameters, such as time steps (Figure \ref{fig:timesteps}) and cell counts (Figure \ref{fig:cell}), linearly shift the data generation baseline without altering the fundamental slope of the cost curve. 

Table \ref{tab:smscl3} quantifies this transition through three representative model tiers, demonstrating the dramatic inversion of the cost profile. For a ``Large'' foundation model ($2\times10^5$ samples), data generation accounts for approximately $96\%$ of the total compute budget (\$$10.3\hspace{1pt}\text{M}$ vs \$$0.4\hspace{1pt}\text{M}$), confirming that for current state-of-the-art surrogates, efficient simulation is the immediate barrier. However, as we scale to the ``XXLarge'' frontier ($2\times10^6$ samples, which also coincides with the independent estimate of the number of samples needed to reasonably cover the entire spectrum of CFD use cases (Appendix \ref{sec:appendixC}) and also leads to a (nominal) overall field error of $1\%$), the regime flips: training costs increase to nearly \$$300\hspace{1pt}\text{M}$, eclipsing the \$$103.7\hspace{1pt}\text{M}$ required for data generation. This analysis underscores a critical strategic evolution for the community; while initial efforts must prioritize efficient data generation, long-term scaling towards a ``universal'' CFD model will ultimately be constrained by the computational demands of model training, necessitating potential architectural innovations.

\begin{table}[htbp]
\centering
\begin{threeparttable}
\caption{Large, XLarge, and XXLarge CFD foundation model estimates.}
\label{tab:smscl3}
\setlength{\tabcolsep}{6pt} 

\begin{tabular}{@{}l c c c@{}}
\toprule
& \textbf{Large} & \textbf{XLarge} & \textbf{XXLarge} \\ 
\midrule
\multicolumn{4}{@{}l}{\textit{Data generation}} \\
Training samples ($N$)          & $2 \times 10^5$      & $1 \times 10^6$      & $2 \times 10^6$ \\
Number of cells ($V$)           & $5 \times 10^8$      & $5 \times 10^8$      & $5 \times 10^8$ \\
Number of steps ($T$)           & $200,000$            & $200,000$            & $200,000$ \\
Flops per cell per step ($f$)       & $2,800$              & $2,800$              & $2,800$ \\
Compute flops ($C$)         & $5.6 \times 10^{22}$ & $2.8 \times 10^{23}$ & $5.6 \times 10^{23}$ \\
GPU hours ($C_{h}$) \tnote{a}  & $1.3 \times 10^{6}$  & $6.48 \times 10^{6}$  & $1.3 \times 10^{7}$ \\
Compute cost ($C_{c}$) \tnote{b} & \$$10.3\hspace{1pt}\text{M}$            & \$$51.8\hspace{1pt}\text{M}$             & \$$103.7\hspace{1pt}\text{M}$ \\
Storage ($S$) \tnote{c}       & $400\hspace{1pt}\text{TB}$                & $2\hspace{1pt}\text{PB}$                  & 4PB \\
Yearly storage cost ($S_c$) \tnote{d}   & \$$48\hspace{1pt}\text{k}$                & \$$240\hspace{1pt}\text{k}$               & \$$480\hspace{1pt}\text{k}$ \\
\midrule
\multicolumn{4}{@{}l}{\textit{Model training}} \\
Data scaling exponent ($\beta$) & $0.43$               & $0.43$               & $0.43$ \\
Effective copies ($K$)          & $50$                 & $50$                 & $50$ \\
Compression ratio ($\delta_K$)  & $5$                  & $5$                  & $5$ \\
Model scaling exponent ($\alpha$) & $0.24$             & $0.24$               & $0.24$ \\
Model size ($M$)                & $3.15 \times 10^9$   & $5.6\times 10^{10}$ & $1.95\times 10^{11}$ \\
Epochs ($n$)                    & $100$                & $100$                & $100$ \\
Flops ($f_{e}$)          & $5.2$              & $5.2$              & $5.2$ \\
Encoding Channel Dimension ($C_{E}$)          & $1024$              & $1024$              & $1024$ \\
Patch Size ($p$)          & $2$              & $2$              & $2$ \\
Number of Latent Tokens ($\tau$) & $32768$ & $32768$  & $32768$ \\
Input size ($I$)                & $4.7 \times 10^{7}$ & $4.7 \times 10^{7}$ & $4.7 \times 10^{7}$ \\
Training flops ($C^T$)          & $5.36\times 10^{23}$ & $4.79 \times 10^{25}$ & $3.99 \times 10^{26}$ \\
GPU hours ($C^{Th}$) \tnote{e}  &  $4.963 \times 10^{4}$                 & $4.43 \times 10^{6}$ & $3.69 \times 10^{7}$ \\
Compute cost ($C^{Tc}$) \tnote{b} & \$$397\hspace{1pt}\text{k}$            & \$$35.5\hspace{1pt}\text{M}$               & \$$295\hspace{1pt}\text{M}$ \\
\bottomrule
\end{tabular}

\begin{tablenotes} \footnotesize
\item [a] Assuming NVIDIA GB200 in FP32 ($80$ TFLOPS) with $15$\% sustained flops usage.
\item [b] Assuming \$$8$ cost per hour for a single NVIDIA GB200 GPU.
\item [c] Assuming $2\hspace{1pt}\text{GB}$ for FP32 with $10$ variables stored on downsampled mesh.
\item [d] Assuming \$$0.01$ per GB per month.
\item [e] Assuming NVIDIA GB200 in BF16 ($5$ PFLOPS) with $60$\% sustained flops usage.
\end{tablenotes}
\end{threeparttable}
\end{table}

\clearpage

\section{Open questions and future work for the community}

We conclude the paper with a number of open-questions for the CFD and ML communities to address.
\paragraph{Reducing the computational expense of data generation} As demonstrated in this paper -- the cost of data generation is a large barrier to the development of a CFD foundation model. 
The main terms that influence this are the number of total samples required, the fidelity of the underlying modeling approach, and the computational efficiency $f$ of the code. 
More rigorous work is needed to understand how to improve the flops per cell per step with a focus on optimizing for latest generation GPUs and consideration of lower precision.

\paragraph{Overcoming the storage bottleneck} As the fidelity of CFD simulations increases, transient simulations, e.g., WMLES and WRLES, are producing massive datasets, frequently exceeding the petabyte threshold. This exponential data volume poses a severe limitation, given that the necessary storage capacity is rapidly outpacing the capabilities of conventional HPC storage architectures. To To address this severe I/O bottleneck, future work must rigorously investigate advanced data compression strategies that go beyond simple subsampling. This includes a re-evaluation of classical dimensionality reduction techniques, such as Proper Orthogonal Decomposition (PoD) to extract dominant flow modes, alongside modern deep learning paradigms like Implicit Neural Representations (Neural Fields) and latent-space compression via autoencoders. Developing these compact, high-fidelity representations is critical to decoupling data generation costs from storage limitations, ultimately enabling the feasible training of large-scale foundation models with transient data. We point out the following surveys that extensively discuss the ``cost of storing data'', and the challenges on increasing permanent storage systems for keeping up with modern day compute capabilities ~\citep{reed_exascale, luttgau_survey, thomas_predicting, hpcwire2025future}. 

\paragraph{Online training} A primary obstacle to developing data-driven models for transient flows is the immense scale of the training data, which can easily reach petabytes. This size makes conventional ``offline'' paradigm of saving data to disk before training computationally infeasible due to I/O bottlenecks and storage costs.

A potential solution is online training, an in-situ co-processing workflow where the machine learning model is trained concurrently with the data-generating CFD simulation. This approach eliminates the need for intermediate storage and inherently allows the model to learn temporal dynamics from time-series data, a critical capability for transient flow modeling. There are however many challenges to this approach, including the vastly different timescales between data generation and model training. 

\paragraph{Domain specific models} The majority of our discussion has been based upon the progression towards a universal model for CFD. A logical question is link to the LLM discussion of domain-specific models rather than frontier-type models with $>100\hspace{1pt}\text{B}$ parameters. The logic being that a small model trained on more specific use-case or single mode dataset may be ultimately cheaper to train and crucially cheaper to inference. A similar logic would be made for CFD. Is there a benefit of allowing the model to see vastly different input space or is there a limit to which this makes sense where small, domain specific models will be better.

\paragraph{Inductive biases} A key open question is in what way inductive biases in the loss function or model architecture can improve performance in the large scale limit. We do not presume to know, and cannot answer this question from the current perspective but see it as an important open question for the community.

\section{Conclusions}
The convergence of high-performance computing and artificial intelligence is fundamentally reshaping the landscape of Computational Fluid Dynamics. In this work, we provided a framework to navigate this new paradigm, aimed at bridging the gap between the machine learning and CFD communities. By deconstructing the industrial CFD workflow into its constituent parts, we have highlighted the vast, input space for CFD.

Central to our contribution is the formulation of a novel scaling law tailored specifically for CFD. Moving beyond direct analogies to large language models, our law introduces critical new dimensions that capture the unique characteristics of scientific simulation. For the first time we also provide estimates on the computational time and cost to develop such foundation models, suggesting it is a tractable problem.

The future of building CFD foundation models is inextricably linked to the development of a new generation of simulation tools: highly efficient, GPU-native CFD solvers. These solvers must be designed to generate vast, high-fidelity datasets quickly, accurately, and at the lowest possible cost, directly addressing the a key term in the compute budget.

In this paper, we have established quantitative estimates for the large-scale
limit, distinguishing between regimes where the cost of data generation is the
dominant factor in total compute versus where the cost of model training prevails.
The coefficients such as $\alpha$ and $\beta$ need to be determined experimentally and likely adapted to different CFD use cases as outlined in the Appendix~\ref{sec:appendixC}.

Crucially, these next-generation solvers cannot be developed in isolation. They must be co-designed for seamless integration with ML frameworks, creating a unified ecosystem where potentially online training can be used. Such integration is the only viable path (in addition to novel compression techniques) to overcoming the immense storage and I/O bottlenecks associated with high-fidelity transient data.

We accept that there are many unanswered questions in the context of CFD foundation models, but we hope this paper can contribute to its advancement.

\begin{ack}
Thank you to the following people for fruitful discussions that helped shape the thoughts shown in this paper: Richard Kurle, Maurits Bleeker, Suchita Kulkarni, William Van Noordt, Rishi Ranade, Daniel Leibovici, Sheel Nidhan, Ira Shokar, Mohammad Amin Nabian, Kaustubh Tangsali, Michael Barad, Peter Sharpe, Jean Kossaifi, Nikola Kovachki, Akshay Subramaniam, Ruben Ohana, Will Trojak.
\end{ack}

\bibliography{mendeley,referencesextra}

@article{Ashton2018g,
    title = {{Assessing the Sensitivity of Hybrid RANS-LES Simulations to Mesh Resolution, Numerical Schemes and Turbulence Modelling within an Industrial CFD Process}},
    year = {2018},
    journal = {SAE Technical Papers},
    author = {Ashton, N. and Unterlechner, P. and Blacha, T.},
    volume = {2018-April},
    doi = {10.4271/2018-01-0709},
    issn = {01487191}
}

@article{Ashton2016,
    title = {{Assessment of RANS and DES methods for realistic automotive models}},
    year = {2016},
    journal = {Computers {\&} Fluids},
    author = {Ashton, N. and West, A. and Lardeau, S. and Revell, A.},
    pages = {1--15},
    volume = {128},
    url = {http://linkinghub.elsevier.com/retrieve/pii/S0045793016000190},
    doi = {10.1016/j.compfluid.2016.01.008},
    issn = {00457930}
}

@article{Khalafvand2011,
    title = {{CFD simulation of flow through heart: a perspective review.}},
    year = {2011},
    journal = {Computer methods in biomechanics and biomedical engineering},
    author = {Khalafvand, S S and Ng, E Y K and Zhong, L},
    number = {1},
    month = {1},
    pages = {113--32},
    volume = {14},
    url = {http://www.ncbi.nlm.nih.gov/pubmed/21271418},
    doi = {10.1080/10255842.2010.493515},
    issn = {1476-8259},
    pmid = {21271418}
}

@article{Slotnick2014,
    title = {{CFD Vision 2030 Study: A Path to Revolutionary Computational Aerosciences}},
    year = {2014},
    journal = {Nasa Cr-2014-21878},
    author = {Slotnick, Jeffrey and Khodadoust, Abdollah and Alonso, Juan and Darmofal, David and Gropp, William and Lurie, Elizabeth and Mavriplis, Dimitri},
    number = {March},
    url = {http://ntrs.nasa.gov/search.jsp?R=20140003093}
}

@article{Spalart2009,
    title = {{Detached-Eddy Simulation}},
    year = {2009},
    journal = {Annual Review of Fluid Mechanics},
    author = {Spalart, Philippe R.},
    number = {1},
    month = {1},
    pages = {181--202},
    volume = {41},
    url = {http://www.annualreviews.org/doi/abs/10.1146/annurev.fluid.010908.165130},
    doi = {10.1146/annurev.fluid.010908.165130},
    issn = {0066-4189}
}

@article{ashton2024drivaer,
    title = {{DrivAerML - High-Fidelity Computational Fluid Dynamics Dataset for Road-Car External Aerodynamics}},
    year = {2024},
    journal = {arxiv.org},
    author = {Ashton, Neil and Mockett, Charles and Fuchs, Marian and Fliessbach, Louis and Hetmann, Hendrik and Knacke, Thilo and Schonwald, Norbert and Skaperdas, Vangelis and Fotiadis, Grigoris and Walle, Astrid and Hupertz, Burkhard and Maddix, Danielle},
    arxivId = {2408.11969}
}

@inproceedings{Heft2014,
    title = {{Experimental and numerical investigation of the drivaer model}},
    year = {2012},
    booktitle = {Proceedings of the ASME 2012 Fluids Engineering Summer meeting},
    author = {Heft, Angelina I and Adams, Nikolaus A},
    pages = {1--11}
}

@article{smagor1963,
    title = {{General circulation experiments with the primitive equations: 1. The basic equations}},
    year = {1963},
    journal = {Mon. Weather Rev.},
    author = {Smagorinsky, J},
    pages = {91--164},
    volume = {91}
}

@article{cetin2023b,
    title = {{HLPW-4: Wall-Modeled Large-Eddy Simulation and Lattice–Boltzmann Technology Focus Group Workshop Summary}},
    year = {2023},
    journal = {Journal of Aircraft},
    author = {Kiris, Cetin C. and Ghate, Aditya S. and Browne, Oliver M. F. and Slotnick, Jeffrey and Larsson, Johan},
    number = {4},
    month = {7},
    pages = {1118--1140},
    volume = {60},
    doi = {10.2514/1.C037193},
    issn = {0021-8669}
}

@article{brunton2019b,
    title = {{Machine Learning for Fluid Mechanics}},
    year = {2019},
    journal = {Annu. Rev. Fluid Mech. 2020},
    author = {Brunton, Steven L and Noack, Bernd R and Koumoutsakos, Petros},
    pages = {477--508},
    volume = {52},
    url = {https://doi.org/10.1146/annurev-fluid-010719-},
    doi = {10.1146/annurev-fluid-010719}
}

@article{Spalart2016,
    title = {{On the role and challenges of CFD in the aerospace industry}},
    year = {2016},
    journal = {The Aeronautical Journal},
    author = {Spalart, P. R. and Venkatakrishnan, V.},
    number = {1223},
    pages = {209--232},
    volume = {120},
    url = {http://www.journals.cambridge.org/abstract_S000192401500010X},
    doi = {10.1017/aer.2015.10},
    issn = {0001-9240}
}

@article{Vermeire2017,
    title = {{On the utility of GPU accelerated high-order methods for unsteady flow simulations: A comparison with industry-standard tools}},
    year = {2017},
    journal = {Journal of Computational Physics},
    author = {Vermeire, B.C. and Witherden, F.D. and Vincent, P.E.},
    pages = {497--521},
    volume = {334},
    publisher = {Elsevier Inc.},
    url = {http://linkinghub.elsevier.com/retrieve/pii/S0021999116307136},
    isbn = {9781624103667},
    doi = {10.1016/j.jcp.2016.12.049},
    issn = {00219991}
}

@inproceedings{Appa2021,
    title = {{Performance of CPU and GPU HPC Architectures for off-design aircraft simulations}},
    year = {2021},
    booktitle = {AIAA Scitech 2021 Forum},
    author = {Appa, Jamil and Turner, Mike and Ashton, Neil},
    number = {January},
    month = {1},
    pages = {11--15},
    publisher = {American Institute of Aeronautics and Astronautics},
    url = {https://arc.aiaa.org/doi/10.2514/6.2021-0141},
    address = {Reston, Virginia},
    isbn = {978-1-62410-609-5},
    doi = {10.2514/6.2021-0141}
}

@article{launder75,
    title = {{Progress in the development of a Reynolds stress turbulence closure}},
    year = {1975},
    journal = {Journal of Fluid Mechanics},
    author = {Launder, B E and Reece, G J and Rodi, W},
    pages = {537--566},
    volume = {68}
}

@article{Witherden2013,
    title = {{PyFR: An open source framework for solving advection-diffusion type problems on streaming architectures using the flux reconstruction approach}},
    year = {2013},
    journal = {Computer Physics Communications},
    author = {Witherden, F. D. and Farrington, a. M. and Vincent, P. E.},
    number = {11},
    pages = {3028--3040},
    volume = {185},
    publisher = {Elsevier B.V.},
    url = {http://dx.doi.org/10.1016/j.cpc.2014.07.011},
    doi = {10.1016/j.cpc.2014.07.011},
    issn = {00104655},
    arxivId = {1312.1638}
}

@article{ashton2023b,
    title = {{Summary of the 4th High-Lift Prediction Workshop Hybrid RANS/LES Technology Focus Group}},
    year = {2023},
    journal = {Journal of Aircraft},
    author = {Ashton, Neil and Batten, Paul and Cary, Andrew and Holst, Kevin},
    month = {8},
    pages = {1--30},
    url = {https://arc.aiaa.org/doi/10.2514/1.C037329},
    doi = {10.2514/1.C037329},
    issn = {0021-8669}
}

@article{Duraisamy2018,
    title = {{Turbulence Modeling in the Age of Data}},
    year = {2018},
    journal = {Annual Review of Fluid Mechanics},
    author = {Duraisamy, Karthik and Iaccarino, Gianluca and Xiao, Heng},
    number = {1},
    pages = {357--377},
    volume = {51},
    doi = {10.1146/annurev-fluid-010518-040547},
    issn = {0066-4189}
}

@article{Menter1994,
    title = {{Two-Equation Eddy-Viscosity Turbulence Models for Engineering Applications}},
    year = {1994},
    journal = {AIAA},
    author = {Menter, F R},
    number = {8},
    pages = {1598--1605},
    volume = {32}
}

@inbook{clark2025HLPW5,
author = {Adam M. Clark and Christopher L. Rumsey and Jeffrey P. Slotnick and Li Wang},
title = {High-Lift Prediction Workshop 5: Overview and Workshop Summary},
booktitle = {AIAA SCITECH 2025 Forum},
doi = {10.2514/6.2025-0045},
URL = {https://arc.aiaa.org/doi/abs/10.2514/6.2025-0045},
eprint = {https://arc.aiaa.org/doi/pdf/10.2514/6.2025-0045},
year={2025}
}

@article{wu2024transolver,
  title={Transolver: A fast transformer solver for pdes on general geometries},
  author={Wu, Haixu and Luo, Huakun and Wang, Haowen and Wang, Jianmin and Long, Mingsheng},
  journal={arXiv preprint arXiv:2402.02366},
  year={2024}
}

@article{li2023geometry,
  title={Geometry-informed neural operator for large-scale 3d pdes},
  author={Li, Zongyi and Kovachki, Nikola and Choy, Chris and Li, Boyi and Kossaifi, Jean and Otta, Shourya and Nabian, Mohammad Amin and Stadler, Maximilian and Hundt, Christian and Azizzadenesheli, Kamyar and others},
  journal={Advances in Neural Information Processing Systems},
  volume={36},
  pages={35836--35854},
  year={2023}
}

@article{ranade2025domino,
  title={DoMINO: A Decomposable Multi-scale Iterative Neural Operator for Modeling Large Scale Engineering Simulations},
  author={Ranade, Rishikesh and Nabian, Mohammad Amin and Tangsali, Kaustubh and Kamenev, Alexey and Hennigh, Oliver and Cherukuri, Ram and Choudhry, Sanjay},
  journal={arXiv preprint arXiv:2501.13350},
  year={2025}
}

@article{alkin2025ab,
  title={AB-UPT: Scaling Neural CFD Surrogates for High-Fidelity Automotive Aerodynamics Simulations via Anchored-Branched Universal Physics Transformers},
  author={Alkin, Benedikt and Bleeker, Maurits and Kurle, Richard and Kronlachner, Tobias and Sonnleitner, Reinhard and Dorfer, Matthias and Brandstetter, Johannes},
  journal={arXiv preprint arXiv:2502.09692},
  year={2025}
}

@article{sharpe2025accelerating,
  title={Accelerating Transient CFD through Machine Learning-Based Flow Initialization},
  author={Sharpe, Peter and Ranade, Rishikesh and Tangsali, Kaustubh and Nabian, Mohammad Amin and Cherukuri, Ram and Choudhry, Sanjay},
  journal={arXiv preprint arXiv:2503.15766},
  year={2025}
}

@book{Launder_Sandham_2002, place={Cambridge}, title={Closure Strategies for Turbulent and Transitional Flows}, author={Launder, Brian and Sandham, Neil }, publisher={Cambridge University Press}, year={2002}}

@misc{siklósi2025reducedmixedprecisionturbulent,
      title={Reduced and mixed precision turbulent flow simulations using explicit finite difference schemes}, 
      author={Bálint Siklósi and Pushpender K. Sharma and David J. Lusher and István Z. Reguly and Neil D. Sandham},
      year={2025},
      eprint={2505.20911},
      archivePrefix={arXiv},
      primaryClass={cs.CE},
      url={https://arxiv.org/abs/2505.20911}, 
}

@article{elrefaie2024drivaernet,
  title={DrivAerNet++: A Large-Scale Multimodal Car Dataset with Computational Fluid Dynamics Simulations and Deep Learning Benchmarks},
  author={Elrefaie, Mohamed and Morar, Florin and Dai, Angela and Ahmed, Faez},
  journal={arXiv preprint arXiv:2406.09624},
  year={2024}
}

@article{lam2023graphcast_new,
      title={Learning skillful medium-range global weather forecasting}, 
      author={Remi Lam and Alvaro Sanchez-Gonzalez and Matthew Willson and Peter Wirnsberger and Meire Fortunato and Ferran Alet and Suman Ravuri and Timo Ewalds and Zach Eaton-Rosen and Weihua Hu and Alexander Merose and Stephan Hoyer and George Holland and Oriol Vinyals and Jacklynn Stott and Alexander Pritzel and Shakir Mohamed and Peter Battaglia},
      year={2023},
      journal={Science},
    volume = {382},
    number = {6677},
    pages = {1416-1421}
}

@inproceedings{lino2023_new,
    title = {{Current and emerging deep-learning methods for the simulation of fluid dynamics}},
    year = {2023},
    booktitle = {Proceedings of the Royal Society A: Mathematical, Physical and Engineering Sciences},
    author = {Lino, Mario and Fotiadis, Stathi and Bharath, Anil A. and Cantwell, Chris D.},
    number = {2275},
    month = {7},
    volume = {479},
    publisher = {Royal Society Publishing},
    doi = {10.1098/rspa.2023.0058},
    issn = {14712946}
}

@article{kaplan2020scaling,
  title={Scaling laws for neural language models},
  author={Kaplan, Jared and McCandlish, Sam and Henighan, Tom and Brown, Tom B and Chess, Benjamin and Child, Rewon and Gray, Scott and Radford, Alec and Wu, Jeffrey and Amodei, Dario},
  journal={arXiv preprint arXiv:2001.08361},
  year={2020}
}

@article{hoffmann2022training,
  title={Training compute-optimal large language models},
  author={Hoffmann, Jordan and Borgeaud, Sebastian and Mensch, Arthur and Buchatskaya, Elena and Cai, Trevor and Rutherford, Eliza and Casas, Diego de Las and Hendricks, Lisa Anne and Welbl, Johannes and Clark, Aidan and others},
  journal={arXiv preprint arXiv:2203.15556},
  year={2022}
}

@article{vaswani2017attention,
  title={Attention is all you need},
  author={Vaswani, Ashish and Shazeer, Noam and Parmar, Niki and Uszkoreit, Jakob and Jones, Llion and Gomez, Aidan N and Kaiser, {\L}ukasz and Polosukhin, Illia},
  journal={Advances in neural information processing systems},
  volume={30},
  year={2017}
}

@article{radford2018improving,
  title={Improving language understanding by generative pre-training},
  author={Radford, Alec and Narasimhan, Karthik and Salimans, Tim and Sutskever, Ilya and others}
}

@inproceedings{zhai2022scaling,
  title={Scaling vision transformers},
  author={Zhai, Xiaohua and Kolesnikov, Alexander and Houlsby, Neil and Beyer, Lucas},
  booktitle={Proceedings of the IEEE/CVF conference on computer vision and pattern recognition},
  pages={12104--12113},
  year={2022}
}

@article{dosovitskiy2020image,
  title={An image is worth 16x16 words: Transformers for image recognition at scale},
  author={Dosovitskiy, Alexey and Beyer, Lucas and Kolesnikov, Alexander and Weissenborn, Dirk and Zhai, Xiaohua and Unterthiner, Thomas and Dehghani, Mostafa and Minderer, Matthias and Heigold, Georg and Gelly, Sylvain and others},
  journal={arXiv preprint arXiv:2010.11929},
  year={2020}
}

@inproceedings{dehghani2023scaling,
  title={Scaling vision transformers to 22 billion parameters},
  author={Dehghani, Mostafa and Djolonga, Josip and Mustafa, Basil and Padlewski, Piotr and Heek, Jonathan and Gilmer, Justin and Steiner, Andreas Peter and Caron, Mathilde and Geirhos, Robert and Alabdulmohsin, Ibrahim and others},
  booktitle={International conference on machine learning},
  pages={7480--7512},
  year={2023},
  organization={PMLR}
}

@article{team2023gemini,
  title={Gemini: a family of highly capable multimodal models},
  author={Team, Gemini and Anil, Rohan and Borgeaud, Sebastian and Alayrac, Jean-Baptiste and Yu, Jiahui and Soricut, Radu and Schalkwyk, Johan and Dai, Andrew M and Hauth, Anja and Millican, Katie and others},
  journal={arXiv preprint arXiv:2312.11805},
  year={2023}
}

@article{achiam2023gpt,
  title={Gpt-4 technical report},
  author={Achiam, Josh and Adler, Steven and Agarwal, Sandhini and Ahmad, Lama and Akkaya, Ilge and Aleman, Florencia Leoni and Almeida, Diogo and Altenschmidt, Janko and Altman, Sam and Anadkat, Shyamal and others},
  journal={arXiv preprint arXiv:2303.08774},
  year={2023}
}

@article{weber2024redpajama,
	title   = {RedPajama: an Open Dataset for Training Large Language Models},
	author  = {Maurice Weber and Daniel Y. Fu and Quentin Anthony and Yonatan Oren and Shane Adams and Anton Alexandrov and Xiaozhong Lyu and Huu Nguyen and Xiaozhe Yao and Virginia Adams and Ben Athiwaratkun and Rahul Chalamala and Kezhen Chen and Max Ryabinin and Tri Dao and Percy Liang and Christopher Ré and Irina Rish and Ce Zhang},
	journal = {NeurIPS Datasets and Benchmarks Track},
	year    = 2024,
}

@article{penedo2024fineweb,
  title={The fineweb datasets: Decanting the web for the finest text data at scale},
  author={Penedo, Guilherme and Kydl{\'\i}{\v{c}}ek, Hynek and Lozhkov, Anton and Mitchell, Margaret and Raffel, Colin A and Von Werra, Leandro and Wolf, Thomas and others},
  journal={Advances in Neural Information Processing Systems},
  volume={37},
  pages={30811--30849},
  year={2024}
}

@article{li2024datacomp,
  title={Datacomp-lm: In search of the next generation of training sets for language models},
  author={Li, Jeffrey and Fang, Alex and Smyrnis, Georgios and Ivgi, Maor and Jordan, Matt and Gadre, Samir Yitzhak and Bansal, Hritik and Guha, Etash and Keh, Sedrick Scott and Arora, Kushal and others},
  journal={Advances in Neural Information Processing Systems},
  volume={37},
  pages={14200--14282},
  year={2024}
}

@article{touvron2023llama,
  title={Llama: Open and efficient foundation language models},
  author={Touvron, Hugo and Lavril, Thibaut and Izacard, Gautier and Martinet, Xavier and Lachaux, Marie-Anne and Lacroix, Timoth{\'e}e and Rozi{\`e}re, Baptiste and Goyal, Naman and Hambro, Eric and Azhar, Faisal and others},
  journal={arXiv preprint arXiv:2302.13971},
  year={2023}
}

@article{grattafiori2024llama,
  title={The llama 3 herd of models},
  author={Grattafiori, Aaron and Dubey, Abhimanyu and Jauhri, Abhinav and Pandey, Abhinav and Kadian, Abhishek and Al-Dahle, Ahmad and Letman, Aiesha and Mathur, Akhil and Schelten, Alan and Vaughan, Alex and others},
  journal={arXiv preprint arXiv:2407.21783},
  year={2024}
}

@book{raffel2018particle,
  title={Particle image velocimetry: a practical guide},
  author={Raffel, Markus and Willert, Christian E and Scarano, Fulvio and K{\"a}hler, Christian J and Wereley, Steve T and Kompenhans, J{\"u}rgen},
  year={2018},
  publisher={springer}
}

@book{adrian2011particle,
  title={Particle image velocimetry},
  author={Adrian, Ronald J and Westerweel, Jerry},
  number={30},
  year={2011},
  publisher={Cambridge university press}
}

@book{forrester2008engineering,
  title={Engineering design via surrogate modelling: a practical guide},
  author={Forrester, Alexander and Sobester, Andras and Keane, Andy},
  year={2008},
  publisher={John Wiley \& Sons}
}

@inproceedings{kipfgcn,
  author       = {Thomas N. Kipf and
                  Max Welling},
  title        = {Semi-Supervised Classification with Graph Convolutional Networks},
  booktitle    = {5th International Conference on Learning Representations, {ICLR} 2017,
                  Toulon, France, April 24-26, 2017, Conference Track Proceedings},
  publisher    = {OpenReview.net},
  year         = {2017}
}

@inproceedings{pfaff2020learning,
  title={Learning mesh-based simulation with graph networks},
  author={Pfaff, Tobias and Fortunato, Meire and Sanchez-Gonzalez, Alvaro and Battaglia, Peter},
  booktitle={International conference on learning representations}
}

@inproceedings{ronneberger2015u,
  title={U-net: Convolutional networks for biomedical image segmentation},
  author={Ronneberger, Olaf and Fischer, Philipp and Brox, Thomas},
  booktitle={International Conference on Medical image computing and computer-assisted intervention},
  pages={234--241},
  year={2015},
  organization={Springer}
}

@article{gupta2022towards,
  title={Towards multi-spatiotemporal-scale generalized pde modeling},
  author={Gupta, Jayesh K and Brandstetter, Johannes},
  journal={arXiv preprint arXiv:2209.15616},
  year={2022}
}

@article{Lu:19,
  title={Deeponet: Learning nonlinear operators for identifying differential equations based on the universal approximation theorem of operators},
  author={Lu, Lu and Jin, Pengzhan and Karniadakis, George Em},
  journal={arXiv preprint arXiv:1910.03193},
  year={2019}
}

@article{Lu:21,
  title={Learning nonlinear operators via DeepONet based on the universal approximation theorem of operators},
  author={Lu, Lu and Jin, Pengzhan and Pang, Guofei and Zhang, Zhongqiang and Karniadakis, George Em},
  journal={Nature machine intelligence},
  volume={3},
  number={3},
  pages={218--229},
  year={2021},
  publisher={Nature Publishing Group UK London}
}

@article{Li:23,
  title={Geometry-Informed Neural Operator for Large-Scale 3D PDEs},
  author={Li, Zongyi and Kovachki, Nikola Borislavov and Choy, Chris and Li, Boyi and Kossaifi, Jean and Otta, Shourya Prakash and Nabian, Mohammad Amin and Stadler, Maximilian and Hundt, Christian and Azizzadenesheli, Kamyar and others},
  journal={arXiv preprint arXiv:2309.00583},
  year={2023}
}

@article{Li:20graph,
  title={Neural operator: Graph kernel network for partial differential equations},
  author={Li, Zongyi and Kovachki, Nikola and Azizzadenesheli, Kamyar and Liu, Burigede and Bhattacharya, Kaushik and Stuart, Andrew and Anandkumar, Anima},
  journal={arXiv preprint arXiv:2003.03485},
  year={2020}
}

@article{Li:20,
  title={Fourier neural operator for parametric partial differential equations},
  author={Li, Zongyi and Kovachki, Nikola and Azizzadenesheli, Kamyar and Liu, Burigede and Bhattacharya, Kaushik and Stuart, Andrew and Anandkumar, Anima},
  journal={arXiv preprint arXiv:2010.08895},
  year={2020}
}

@article{Kovachki:21,
  title={Neural operator: Learning maps between function spaces},
  author={Kovachki, Nikola and Li, Zongyi and Liu, Burigede and Azizzadenesheli, Kamyar and Bhattacharya, Kaushik and Stuart, Andrew and Anandkumar, Anima},
  journal={arXiv preprint arXiv:2108.08481},
  year={2021}
}

@article{bi_accurate_2023,
  author       = {Kaifeng Bi and
                  Lingxi Xie and
                  Hengheng Zhang and
                  Xin Chen and
                  Xiaotao Gu and
                  Qi Tian},
  title        = {Accurate medium-range global weather forecasting with 3D neural networks},
  journal      = {Nat.},
  volume       = {619},
  number       = {7970},
  pages        = {533--538},
  year         = {2023},
  doi          = {10.1038/S41586-023-06185-3},
  bibsource    = {dblp computer science bibliography, https://dblp.org}
}

@article{bodnar_aurora_2024,
  author       = {Cristian Bodnar and
                  Wessel P. Bruinsma and
                  Ana Lucic and
                  Megan Stanley and
                  Johannes Brandstetter and
                  Patrick Garvan and
                  Maik Riechert and
                  Jonathan A. Weyn and
                  Haiyu Dong and
                  Anna Vaughan and
                  Jayesh K. Gupta and
                  Kit Thambiratnam and
                  Alex Archibald and
                  Elizabeth Heider and
                  Max Welling and
                  Richard E. Turner and
                  Paris Perdikaris},
  title        = {Aurora: {A} Foundation Model of the Atmosphere},
  journal      = {CoRR},
  volume       = {abs/2405.13063},
  year         = {2024},
  doi          = {10.48550/ARXIV.2405.13063},
  eprinttype    = {arXiv},
  eprint       = {2405.13063},
  bibsource    = {dblp computer science bibliography, https://dblp.org}
}

@article{jumper_highly_2021,
  title={Highly accurate protein structure prediction with AlphaFold},
  author={Jumper, John and Evans, Richard and Pritzel, Alexander and Green, Tim and Figurnov, Michael and Ronneberger, Olaf and Tunyasuvunakool, Kathryn and Bates, Russ and {\v{Z}}{\'\i}dek, Augustin and Potapenko, Anna and others},
  journal={nature},
  volume={596},
  number={7873},
  pages={583--589},
  year={2021},
  publisher={Nature Publishing Group}
}

@article{abramson_accurate_2024,
  title={Accurate structure prediction of biomolecular interactions with AlphaFold 3},
  author={Abramson, Josh and Adler, Jonas and Dunger, Jack and Evans, Richard and Green, Tim and Pritzel, Alexander and Ronneberger, Olaf and Willmore, Lindsay and Ballard, Andrew J and Bambrick, Joshua and others},
  journal={Nature},
  pages={1--3},
  year={2024},
  publisher={Nature Publishing Group UK London}
}

@article{li2022transformer,
  title={Transformer for partial differential equations' operator learning},
  author={Li, Zijie and Meidani, Kazem and Farimani, Amir Barati},
  journal={arXiv preprint arXiv:2205.13671},
  year={2022}
}

@article{alkin2024universal,
  title={Universal physics transformers: A framework for efficiently scaling neural operators},
  author={Alkin, Benedikt and F{\"u}rst, Andreas and Schmid, Simon and Gruber, Lukas and Holzleitner, Markus and Brandstetter, Johannes},
  journal={Advances in Neural Information Processing Systems},
  volume={37},
  pages={25152--25194},
  year={2024}
}

@inproceedings{jaegle2021perceiver,
  title={Perceiver: General perception with iterative attention},
  author={Jaegle, Andrew and Gimeno, Felix and Brock, Andy and Vinyals, Oriol and Zisserman, Andrew and Carreira, Joao},
  booktitle={International conference on machine learning},
  pages={4651--4664},
  year={2021},
  organization={PMLR}
}

@inproceedings{jaegle2021perceiverIO,
  title={Perceiver IO: A General Architecture for Structured Inputs \& Outputs},
  author={Jaegle, Andrew and Borgeaud, Sebastian and Alayrac, Jean-Baptiste and Doersch, Carl and Ionescu, Catalin and Ding, David and Koppula, Skanda and Zoran, Daniel and Brock, Andrew and Shelhamer, Evan and others},
  booktitle={International Conference on Learning Representations},
  year={2021}
}

@article{raissi2019physics,
  title={Physics-informed neural networks: A deep learning framework for solving forward and inverse problems involving nonlinear partial differential equations},
  author={Raissi, Maziar and Perdikaris, Paris and Karniadakis, George E.},
  journal={Journal of Computational physics},
  volume={378},
  pages={686--707},
  year={2019},
  publisher={Elsevier}
}

@article{merchant_scaling_2023,
  author       = {Amil Merchant and
                  Simon L. Batzner and
                  Samuel S. Schoenholz and
                  Muratahan Aykol and
                  Gowoon Cheon and
                  Ekin Dogus Cubuk},
  title        = {Scaling deep learning for materials discovery},
  journal      = {Nat.},
  volume       = {624},
  number       = {7990},
  pages        = {80--85},
  year         = {2023},
  doi          = {10.1038/S41586-023-06735-9},
  bibsource    = {dblp computer science bibliography, https://dblp.org}
}

@book{piegl2012nurbs,
  title={The NURBS Book},
  author={Piegl, L. and Tiller, W.},
  isbn={9783642973857},
  series={Monographs in Visual Communication},
  url={https://books.google.co.uk/books?id=CkeqCAAAQBAJ},
  year={2012},
  publisher={Springer Berlin Heidelberg}
}

@article{coons67,
author = {Coons, Steven},
year = {1967},
month = {06},
pages = {105},
title = {Surfaces for Computer Aided Design of Space Forms},
volume = {41},
journal = {MIT, MAC-TR}
}

@article{seo2005,
author = {Seo, Joonho and Youngjae, Song and Kim, Sungchan and Lee, Kunwoo and Choi, Young and Chae, Soowon},
year = {2005},
month = {01},
pages = {67-76},
title = {Wrap-around Operation for Multi-resolution CAD Model},
volume = {2},
journal = {Computer-Aided Design \& Applications},
doi = {10.1080/16864360.2005.10738354}
}

@article{zeni_generative_2025,
  title={A generative model for inorganic materials design},
  author={Zeni, Claudio and Pinsler, Robert and Z{\"u}gner, Daniel and Fowler, Andrew and Horton, Matthew and Fu, Xiang and Wang, Zilong and Shysheya, Aliaksandra and Crabb{\'e}, Jonathan and Ueda, Shoko and others},
  journal={Nature},
  pages={1--3},
  year={2025},
  publisher={Nature Publishing Group UK London}
}

@article{yang_mattersim_2024,
      title={MatterSim: A Deep Learning Atomistic Model Across Elements, Temperatures and Pressures},
      author={Han Yang and Chenxi Hu and Yichi Zhou and Xixian Liu and Yu Shi and Jielan Li and Guanzhi Li and Zekun Chen and Shuizhou Chen and Claudio Zeni and Matthew Horton and Robert Pinsler and Andrew Fowler and Daniel Zügner and Tian Xie and Jake Smith and Lixin Sun and Qian Wang and Lingyu Kong and Chang Liu and Hongxia Hao and Ziheng Lu},
      year={2024},
      eprint={2405.04967},
      archivePrefix={arXiv},
      primaryClass={cond-mat.mtrl-sci},
      journal={arXiv preprint arXiv:2405.04967}
}

@inproceedings{kirillov2023segment,
  title={Segment anything},
  author={Kirillov, Alexander and Mintun, Eric and Ravi, Nikhila and Mao, Hanzi and Rolland, Chloe and Gustafson, Laura and Xiao, Tete and Whitehead, Spencer and Berg, Alexander C and Lo, Wan-Yen and others},
  booktitle={Proceedings of the IEEE/CVF international conference on computer vision},
  pages={4015--4026},
  year={2023}
}

@inproceedings{rombach2022high,
  title={High-resolution image synthesis with latent diffusion models},
  author={Rombach, Robin and Blattmann, Andreas and Lorenz, Dominik and Esser, Patrick and Ommer, Bj{\"o}rn},
  booktitle={Proceedings of the IEEE/CVF conference on computer vision and pattern recognition},
  pages={10684--10695},
  year={2022}
}

@article{sanderse2024scientific,
  title={Scientific machine learning for closure models in multiscale problems: A review},
  author={Sanderse, Benjamin and Stinis, Panos and Maulik, Romit and Ahmed, Shady E},
  journal={arXiv preprint arXiv:2403.02913},
  year={2024}
}

@article{singh2016using,
  title={Using field inversion to quantify functional errors in turbulence closures},
  author={Singh, Anand Pratap and Duraisamy, Karthik},
  journal={Physics of Fluids},
  volume={28},
  number={4},
  year={2016},
  publisher={AIP Publishing}
}

@article{wang2017physics,
  title={Physics-informed machine learning approach for reconstructing Reynolds stress modeling discrepancies based on DNS data},
  author={Wang, Jian-Xun and Wu, Jin-Long and Xiao, Heng},
  journal={Physical Review Fluids},
  volume={2},
  number={3},
  pages={034603},
  year={2017},
  publisher={APS}
}

@article{gamahara2017searching,
  title={Searching for turbulence models by artificial neural network},
  author={Gamahara, Masataka and Hattori, Yuji},
  journal={Physical Review Fluids},
  volume={2},
  number={5},
  pages={054604},
  year={2017},
  publisher={APS}
}

@article{maulik2019subgrid,
  title={Subgrid modelling for two-dimensional turbulence using neural networks},
  author={Maulik, Romit and San, Omer and Rasheed, Adil and Vedula, Prakash},
  journal={Journal of Fluid Mechanics},
  volume={858},
  pages={122--144},
  year={2019},
  publisher={Cambridge University Press}
}

@article{luo2025transolver++,
  title={Transolver++: An Accurate Neural Solver for PDEs on Million-Scale Geometries},
  author={Luo, Huakun and Wu, Haixu and Zhou, Hang and Xing, Lanxiang and Di, Yichen and Wang, Jianmin and Long, Mingsheng},
  journal={arXiv preprint arXiv:2502.02414},
  year={2025}
}

@article{reed_exascale,
author = {Reed, Daniel A. and Dongarra, Jack},
title = {Exascale computing and big data},
year = {2015},
issue_date = {July 2015},
publisher = {Association for Computing Machinery},
address = {New York, NY, USA},
volume = {58},
number = {7},
issn = {0001-0782},
url = {https://doi.org/10.1145/2699414},
doi = {10.1145/2699414},
journal = {Commun. ACM},
month = jun,
pages = {56–68},
numpages = {13}
}

@article{luttgau_survey,
author = {Luttgau, Jakob and Kuhn, Michael and Duwe, Kira and Alforov, Yevhen and Betke, Eugen and Kunkel, Julian and Ludwig, Thomas},
title = {Survey of Storage Systems for High-Performance Computing},
year = {2018},
issue_date = {March 2018},
publisher = {South Ural State University},
address = {Chelyabinsk, RUS},
volume = {5},
number = {1},
issn = {2409-6008},
url = {https://doi.org/10.14529/jsfi180103},
doi = {10.14529/jsfi180103},
journal = {Supercomput. Front. Innov.: Int. J.},
month = mar,
pages = {31–58},
numpages = {28}
}

@inproceedings{thomas_predicting,
author = {Thomas, Luis and Gougeaud, Sebastien and Rubini, St\'{e}phane and Deniel, Philippe and Boukhobza, Jalil},
title = {Predicting file lifetimes for data placement in multi-tiered storage systems for HPC},
year = {2021},
isbn = {9781450383028},
publisher = {Association for Computing Machinery},
address = {New York, NY, USA},
url = {https://doi.org/10.1145/3439839.3458733},
doi = {10.1145/3439839.3458733},
booktitle = {Proceedings of the Workshop on Challenges and Opportunities of Efficient and Performant Storage Systems},
articleno = {2},
numpages = {9},
location = {Online Event, United Kingdom},
series = {CHEOPS '21}
}

@misc{hpcwire2025future,
  author       = {HPCwire},
  title        = {Future of Storage: HPC and AI Storage by the Numbers},
  howpublished = {\url{https://www.hpcwire.com/2025/10/16/future-of-storage-hpc-and-ai-storage-by-the-numbers/}},
  year         = {2025},
  month        = oct,
  note         = {Accessed: 2025-11-13}
}

@inproceedings{turner2021,
	title = {Performance of cpu and gpu hpc architectures for off-design aircraft simulation},
	isbn = {978-1-62410-609-5},
	booktitle = {{AIAA} {Scitech} 2021 {Forum}},
	author = {Turner, M. and Appa, J. and Ashton, N.},
	year = {2021},
}

@inproceedings{sennrich2016neural,
  title={Neural machine translation of rare words with subword units},
  author={Sennrich, Rico and Haddow, Barry and Birch, Alexandra},
  booktitle={Proceedings of the 54th annual meeting of the association for computational linguistics (volume 1: long papers)},
  pages={1715--1725},
  year={2016}
}

@article{wen2025geometry,
  title={Geometry aware operator transformer as an efficient and accurate neural surrogate for pdes on arbitrary domains},
  author={Wen, Shizheng and Kumbhat, Arsh and Lingsch, Levi and Mousavi, Sepehr and Zhao, Yizhou and Chandrashekar, Praveen and Mishra, Siddhartha},
  journal={arXiv preprint arXiv:2505.18781},
  year={2025}
}

@article{Larsson2015v2,
	title = {Large eddy simulation with modeled wall-stress: recent progress and future directions},
	volume = {3},
	doi = {10.1299/mer.15-00418},
	number = {1},
	journal = {Mechanical Engineering Reviews},
	author = {Larsson, Johan and Kawai, Soshi and Bodart, Julien and Bermejo-Moreno, Ivan},
	year = {2015},
	note = {ISBN: 0123456789},
	pages = {15--00418--15--00418},
}

@article{herde2024poseidon,
  title={Poseidon: Efficient foundation models for pdes},
  author={Herde, Maximilian and Raonic, Bogdan and Rohner, Tobias and K{\"a}ppeli, Roger and Molinaro, Roberto and de B{\'e}zenac, Emmanuel and Mishra, Siddhartha},
  journal={Advances in Neural Information Processing Systems},
  volume={37},
  pages={72525--72624},
  year={2024}
}

@article{zhen2023,
    author = {Li, Zhen and Ju, Yaping and Zhang, Chuhua},
    title = {Machine-learning data-driven modeling of laminar-turbulent transition in compressor cascade},
    journal = {Physics of Fluids},
    volume = {35},
    number = {8},
    pages = {085133},
    year = {2023},
    month = {08},
    issn = {1070-6631},
    doi = {10.1063/5.0164131},
    url = {https://doi.org/10.1063/5.0164131},
    eprint = {https://pubs.aip.org/aip/pof/article-pdf/doi/10.1063/5.0164131/18096504/085133_1_5.0164131.pdf},
}

@inbook{nielsen2024,
author = {Eric J. Nielsen and Aaron Walden and Gabriel Nastac and Li Wang and William Jones and Mark Lohry and William K. Anderson and Boris Diskin and Yi Liu and Christopher L. Rumsey and Prahladh Iyer and Patrick Moran and Mohammad Zubair},
title = {Large-Scale Computational Fluid Dynamics Simulations of Aerospace Configurations on the Frontier Exascale System},
booktitle = {AIAA AVIATION FORUM AND ASCEND 2024},
chapter = {},
pages = {},
doi = {10.2514/6.2024-3866},
URL = {https://arc.aiaa.org/doi/abs/10.2514/6.2024-3866},
eprint = {https://arc.aiaa.org/doi/pdf/10.2514/6.2024-3866}
}

@inproceedings{hosseinverdi2025rapidus,
  title={Rapidus: Performance-Portable Parallel Flow Solver for Aerospace Applications},
  author={Hosseinverdi, Shirzad and Sitaraman, Jay and Jude, Dylan and Premaratne, Pavithra and Erlandson, Lucas and Appelhans, David},
  booktitle={AIAA SCITECH 2025 Forum},
  pages={1484},
  year={2025}
}

@article{monaghan92,
  title={Smoothed particle hydrodynamics},
  author={Monaghan, Joe J},
  journal={Annual Review of Astronomy and Astrophysics},
  volume={30},
  number={1},
  pages={543--574},
  year={1992},
  publisher={Annual Reviews}
}

@article{chen98,
  title={Lattice Boltzmann method for fluid flows},
  author={Chen, Shiyi and Doolen, Gary D},
  journal={Annual Review of Fluid Mechanics},
  volume={30},
  number={1},
  pages={329--364},
  year={1998},
  publisher={Annual Reviews}
}

@misc{tompson2022acceleratingeulerianfluidsimulation,
      title={Accelerating Eulerian Fluid Simulation With Convolutional Networks}, 
      author={Jonathan Tompson and Kristofer Schlachter and Pablo Sprechmann and Ken Perlin},
      year={2022},
      eprint={1607.03597},
      archivePrefix={arXiv},
      primaryClass={cs.CV},
      url={https://arxiv.org/abs/1607.03597}, 
}

@inproceedings{guo2016,
author = {Guo, Xiaoxiao and Li, Wei and Iorio, Francesco},
title = {Convolutional Neural Networks for Steady Flow Approximation},
year = {2016},
isbn = {9781450342322},
publisher = {Association for Computing Machinery},
address = {New York, NY, USA},
url = {https://doi.org/10.1145/2939672.2939738},
doi = {10.1145/2939672.2939738},
booktitle = {Proceedings of the 22nd ACM SIGKDD International Conference on Knowledge Discovery and Data Mining},
pages = {481–490},
numpages = {10},
location = {San Francisco, California, USA},
series = {KDD '16}
}

@misc{nie2025largelanguagediffusionmodels,
      title={Large Language Diffusion Models}, 
      author={Shen Nie and Fengqi Zhu and Zebin You and Xiaolu Zhang and Jingyang Ou and Jun Hu and Jun Zhou and Yankai Lin and Ji-Rong Wen and Chongxuan Li},
      year={2025},
      eprint={2502.09992},
      archivePrefix={arXiv},
      primaryClass={cs.CL},
      url={https://arxiv.org/abs/2502.09992}, 
}

@article{BREHM2015184,
title = {A comparison of higher-order finite-difference shock capturing schemes},
journal = {Computers \& Fluids},
volume = {122},
pages = {184-208},
year = {2015},
issn = {0045-7930},
doi = {https://doi.org/10.1016/j.compfluid.2015.08.023},
url = {https://www.sciencedirect.com/science/article/pii/S0045793015002996},
author = {Christoph Brehm and Michael F. Barad and Jeffrey A. Housman and Cetin C. Kiris}
}

@article{williams2009,
author = {Williams, Samuel and Waterman, Andrew and Patterson, David},
title = {Roofline: an insightful visual performance model for multicore architectures},
year = {2009},
issue_date = {April 2009},
publisher = {Association for Computing Machinery},
address = {New York, NY, USA},
volume = {52},
number = {4},
issn = {0001-0782},
url = {https://doi.org/10.1145/1498765.1498785},
doi = {10.1145/1498765.1498785},
journal = {Commun. ACM},
month = apr,
pages = {65–76},
numpages = {12}
}

@article{brandstetter2022message,
  title={Message passing neural PDE solvers},
  author={Brandstetter, Johannes and Worrall, Daniel and Welling, Max},
  journal={arXiv preprint arXiv:2202.03376},
  year={2022}
}

@article{LMK1,
    title = "Error estimates for {D}eep{ON}ets: {A} deep learning framework in infinite dimensions",
    journal = "Transactions of Mathematics and Its Applications",
    volume = "6",
    number = "1",
    pages = "tnac001",
    year = "2022",
    publisher = "Oxford University Press",
    eprint = "https://academic.oup.com/imatrm/article-pdf/6/1/tnac001/42785544/tnac001.pdf",
    author = "Lanthaler, Samuel  and Mishra, Siddhartha  and Karniadakis, George E"
}

@misc{deryck2022genericboundsapproximationerror,
      title={Generic bounds on the approximation error for physics-informed (and) operator learning}, 
      author={Tim De Ryck and Siddhartha Mishra},
      year={2022},
      eprint={2205.11393},
      archivePrefix={arXiv},
      primaryClass={cs.LG},
      url={https://arxiv.org/abs/2205.11393}, 
}

@misc{walrus,
      title={Walrus: A Cross-Domain Foundation Model for Continuum Dynamics}, 
      author={Michael McCabe and Payel Mukhopadhyay and Tanya Marwah and Bruno Regaldo-Saint Blancard and Francois Rozet and Cristiana Diaconu and Lucas Meyer and Kaze W. K. Wong and Hadi Sotoudeh and Alberto Bietti and Irina Espejo and Rio Fear and Siavash Golkar and Tom Hehir and Keiya Hirashima and Geraud Krawezik and Francois Lanusse and Rudy Morel and Ruben Ohana and Liam Parker and Mariel Pettee and Jeff Shen and Kyunghyun Cho and Miles Cranmer and Shirley Ho},
      year={2025},
      eprint={2511.15684},
      archivePrefix={arXiv},
      primaryClass={cs.LG},
      url={https://arxiv.org/abs/2511.15684}, 
}

@misc{mpp,
title={Multiple Physics Pretraining for Physical Surrogate Models},
author={Michael McCabe and Bruno R{\'e}galdo-Saint Blancard and Liam Holden Parker and Ruben Ohana and Miles Cranmer and Alberto Bietti and Michael Eickenberg and Siavash Golkar and Geraud Krawezik and Francois Lanusse and Mariel Pettee and Tiberiu Tesileanu and Kyunghyun Cho and Shirley Ho},
year={2024},
url={https://openreview.net/forum?id=fH9eqpCcR3}
}

@inproceedings{dpot,
title={{DPOT}: Auto-Regressive Denoising Operator Transformer for Large-Scale {PDE} Pre-Training},
author={Zhongkai Hao and Chang Su and Songming Liu and Julius Berner and Chengyang Ying and Hang Su and Anima Anandkumar and Jian Song and Jun Zhu},
booktitle={Forty-first International Conference on Machine Learning},
year={2024},
url={https://openreview.net/forum?id=X7UnDevHOM}
}

@misc{rigno,
      title={RIGNO: A Graph-based framework for robust and accurate operator learning for PDEs on arbitrary domains}, 
      author={Sepehr Mousavi and Shizheng Wen and Levi Lingsch and Maximilian Herde and Bogdan Raonić and Siddhartha Mishra},
      year={2025},
      eprint={2501.19205},
      archivePrefix={arXiv},
      primaryClass={cs.LG},
      url={https://arxiv.org/abs/2501.19205}, 
}

@inproceedings{
CNO,
title={Convolutional Neural Operators for robust and accurate learning of {PDE}s},
author={Bogdan Raonic and Roberto Molinaro and Tim De Ryck and Tobias Rohner and Francesca Bartolucci and Rima Alaifari and Siddhartha Mishra and Emmanuel de Bezenac},
booktitle={Thirty-seventh Conference on Neural Information Processing Systems},
year={2023},
url={https://openreview.net/forum?id=MtekhXRP4h}
}

@inproceedings{
bartolucci2023representation,
title={Representation Equivalent Neural Operators: a Framework for Alias-free Operator Learning},
author={Francesca Bartolucci and Emmanuel de Bezenac and Bogdan Raonic and Roberto Molinaro and Siddhartha Mishra and Rima Alaifari},
booktitle={Thirty-seventh Conference on Neural Information Processing Systems},
year={2023},
url={https://openreview.net/forum?id=7LSEkvEGCM}
}

@inproceedings{liu2021swin,
  title={Swin transformer: Hierarchical vision transformer using shifted windows},
  author={Liu, Ze and Lin, Yutong and Cao, Yue and Hu, Han and Wei, Yixuan and Zhang, Zheng and Lin, Stephen and Guo, Baining},
  booktitle={Proceedings of the IEEE/CVF international conference on computer vision},
  pages={10012--10022},
  year={2021}
}
\bibliographystyle{neurips_2025}


\appendix
\addtocontents{toc}{\protect\setcounter{tocdepth}{-1}}

\section{Further discussion on the role of AI in CFD}
\label{sec:appendixA}
The main text of the paper discusses largely the use of AI surrogates outside of a conventional CFD solver. There are additional areas where AI can help to accelerate a convention CFD process, which we now discuss:

\begin{enumerate}
    \item AI to develop improved turbulence/transition/chemistry models that plug within a conventional CFD solver
    \item AI to precondition/initialize CFD solutions to speed up the time to reach a converged/time-averaged solution using an AI surrogate model.
\end{enumerate}

\subsection{Improved turbulence/transition/chemistry models as plugin for conventional CFD solver}

A significant application of AI in computational fluid dynamics (CFD) lies in the development and calibration of closure models for physical phenomena that are not fully resolved on the computational grid~\citep{brunton2019b, sanderse2024scientific,Duraisamy2018}. In many practical engineering simulations, methods like Direct Numerical Simulation (DNS), which resolve all spatio-temporal scales of turbulence, are computationally intractable. Consequently, methodologies such as Reynolds-Averaged Navier-Stokes (RANS) and Large Eddy Simulation (LES) are employed, which rely on modeling the effects of the unresolved scales. AI and machine learning present a potential alternative to constructing these models with better accuracy.

\subsubsection{Data-driven turbulence models}

The primary challenge in RANS and LES is the formulation of a closure for the unclosed terms that arise from the averaging or filtering process. As shown in Equation \ref{eq:rans1},
many RANS models rely on the Boussinesq hypothesis, which assumes a linear relationship between the Reynolds stress and the mean strain-rate tensor.
Here, the turbulent viscosity, $\mu_t$, is determined by solving additional transport equations (e.g., for turbulent kinetic energy $k$ and dissipation rate $\epsilon$). The core idea of an AI-driven approach is to learn a superior functional mapping, from high-fidelity data (e.g., DNS or experiments) to predict the Reynolds stress tensor. This can be approached by correcting the turbulent viscosity: The idea is to learn a corrective function or a direct mapping for $\mu_t = f (S_{ij}, \Omega_{ij}, k, \epsilon, \dots)$~\citep{singh2016using,wang2017physics}, where $S_{ij}$ is the mean strain rate tensor, and $\Omega_{ij}$ is the mean rotation rate tensor.    
The objective is to create a RANS-based model whose solution field, achieves an accuracy comparable to LES but with a computational cost for RANS that remains orders of magnitude lower than that of LES.

In LES, the governing equations are spatially filtered, introducing the unclosed sub-grid scale (SGS) stress tensor as shown in Equation \ref{eq:sgs_model}. AI models~\citep{gamahara2017searching,maulik2019subgrid} can learn a mapping from the  velocity field to the SGS stress tensor, $\mu_{sgs}$.

\subsubsection{Other AI-enhanced closure models}

The closure paradigm extends beyond turbulence modeling. Laminar-to-turbulent transition can be modeled by learning an intermittency factor \citep{zhen2023}, $\gamma(\vec{x}, t) \in [0, 1]$. An AI model can learn the mapping $\gamma = f(\text{Re}, \nabla p, \dots)$ from local flow parameters to this factor, which then modulates the production terms in a turbulence model like $k$-$\omega$ SST.
In reacting flows (e.g., combustion), the species transport equations contain highly non-linear chemical source terms, which are computationally expensive to evaluate directly using Arrhenius kinetics. An AI model can act as a surrogate, learning the complex mapping from the local thermochemical state (temperature $T$, pressure $P$, species mass fractions $Y_k$) to the source terms, which could replace costly integration or cumbersome lookup tables.

\subsection{AI to precondition/initialize CFD solutions}

A compelling application for AI is the use of surrogate models to provide high-quality initial conditions for iterative CFD solvers, thereby accelerating convergence to a steady-state or target solution \citep{sharpe2025accelerating}.

The central hypothesis is that the surrogate's prediction lies significantly closer in the solution space to the final converged state than a conventional initialization, 
By initializing the solver within a stronger basin of attraction for the true solution, the number of iterations required for convergence can be substantially reduced. Such AI-based method offer a sophisticated alternative to conventional initialization strategies, which typically include either initialization with uniform freestream conditions, or interpolating a solution from a coarser computational mesh or a lower-fidelity simulation (e.g., using a steady-state RANS solution to initialize a scale-resolving simulation).
\section{Estimating flops per cell per step}
\label{sec:appendixB}

This section provides a computational cost analysis, in floating-point operations (FLOPs), for two prevalent CFD methodologies. The objective is to derive the number of FLOPs per computational cell per step ($f$) to estimate resources for large-scale simulations. We accept that there exists a broad range of CFD input combinations, each resulting in different values for $f$. Nevertheless we pick two that correspond to the analysis in this paper and represent a representative high-fidelity and low-fidelity CFD input combination. 

The two methodologies are:
\begin{enumerate}
    \item Implicit steady-state RANS: An unstructured finite-volume method with an implicit solver and the k-$\omega$ SST turbulence model.
    \item Explicit WMLES: A time-resolved Cartesian grid method using a Jameson-Schmidt-Turkel (JST) scheme and the WALE subgrid-scale model.
\end{enumerate}

The derived $f$ metric shows how the two methods diverge, as summarized in Table \ref{table:flops}.

\begin{table}[h!]
\centering
\caption{Summary of the $f$ parameter (FP32).}
\label{table:flops}
\begin{tabular}{|p{4cm}|p{4cm}|p{4cm}|}
\hline
\textbf{CFD methodology} & \textbf{$f$ (FLOPs per cell perstep)} & \textbf{Key cost driver} \\
\hline
Implicit steady-state RANS & \textasciitilde $40.000$ (Range: $20\hspace{1pt}\text{k}-60\hspace{1pt}\text{k}$) & Linear system solution (solver + preconditioner efficiency) \\
\hline
Explicit WMLES & \textasciitilde $2.800$ (Range: $1.8\hspace{1pt}\text{k}-3\hspace{1pt}\text{k}$) & Number of time steps (dictated by CFL stability limit) \\
\hline
\end{tabular}
\end{table}

The wide range for implicit RANS is due to its cost being dominated by the iterative solution of a large, sparse linear system; its performance is highly sensitive to the preconditioner and number of inner solver iterations. In contrast, the explicit LES cost per step is deterministic and low, but the total simulation cost is driven by the vast number of time steps mandated by strict CFL stability constraints.

\subsection{The unified cost formula}

The total computational cost of a CFD simulation can be estimated using the following unified formula:

\begin{equation}
    \text{Total cost (FLOPs)} = V \times T \times f \ ,
\end{equation}

where $V$ is the total cell count, $T$ is the total number of iterations (RANS) or time steps (LES), and $f$ is the methodology-specific parameter from Table \ref{table:flops}.

\subsection{Implicit vs. explicit solvers}
The cost-per-cell metric diverges due to solver architecture. 
\begin{itemize}
    \item Implicit solvers (usually typically for RANS) perform a few computationally expensive steps, dominated by solving a large, coupled linear system, where the state at $t^{n+1}$ depends on all neighbors at $t^{n+1}$.
    \item Explicit solvers (typically used for LES) perform many computationally cheap steps, as the state at $t^{n+1}$ is calculated using only known values from $t^n$. This approach is limited by the strict Courant-Friedrichs-Lewy (CFL) condition, mandating a vast number of small time steps.
\end{itemize}
A true comparison of work requires multiplying $f$ by the total $T$, which will differ by orders of magnitude between the two approaches.

\subsection{Cost analysis of a steady-state implicit RANS solver}

An implicit RANS iteration solves the nonlinear system $R(w) = 0$ using a Newton-like method. A single iteration involves three main stages:
\begin{enumerate}
    \item Residual calculation: Evaluation of the current flux imbalance, $R(w^k)$. The residual $R(w^k)$ represents the net imbalance of fluxes (mass, momentum, and energy) in each control volume of the computational domain, based on the current approximate solution vector $w^k$. If the system were perfectly solved (at steady-state), the residual would be zero. The calculation involves discretizing the RANS equations, which means computing all convective, diffusive, and source terms using the current flow variables ($w^k$). For the 5-equation mean flow (using Roe's scheme) and the 2-equation k-$\omega$ SST model, this involves complex flux/source term evaluation. The combined cost is estimated at \textasciitilde$6\hspace{1pt}\text{k}$ FLOPs/cell.
    \item Jacobian assembly: Calculating and populating the entries of a large, sparse Jacobian $\frac{\partial R}{\partial w}$ during each iteration. A common rule of thumb is that this costs $2-4\times$ the residual calculation. We estimate \textasciitilde$12\hspace{1pt}\text{k}$ FLOPs/cell.
    \item Linear system solution: Solving the large, sparse, linear system for the solution update, $\Delta w$. A block ILU preconditioner is required for the Krylov solver. Applying this involves forward/backward substitution, estimated at \textasciitilde$1.5\hspace{1pt}\text{k}$ FLOPs per cell per inner iteration. The GMRES iterative solver's cost depends on $N_{\text{GMRES}}$. Each iteration involves a sparse matrix-vector product (SpMV) (\textasciitilde700 FLOPs), preconditioner application (\textasciitilde$1.5\hspace{1pt}\text{k}$ FLOPs), and vector operations (\textasciitilde420 FLOPs), for a total of \textasciitilde$2.6\hspace{1pt}\text{k}$ FLOPs per cell per GMRES iteration.
    
\end{enumerate}
Of these, the linear system solution consumes 80\% or more of the total time.

\subsubsection{Synthesis and total FLOP count for RANS}
The total cost per outer RANS iteration sums these components. Assuming a typical $N_{\text{GMRES}} = 10$:
\begin{itemize}
    \item Linear solve cost: $10 \times 2.6\hspace{1pt}\text{k} \approx 26.2\hspace{1pt}\text{k}$ FLOPs per cell.
    \item Total iteration cost: $6\hspace{1pt}\text{k}$ (Residual) + $12\hspace{1pt}\text{k}$ (Jacobian) + $26.2\hspace{1pt}\text{k}$ (Linear solve) $\approx$ $44.2\hspace{1pt}\text{k}$ FLOPs per cell per iteration.
\end{itemize}
This aligns with our summary table and highlights the dominance of the linear solve.

\begin{table}[h!]
\centering
\caption{Estimated FLOPs breakdown per cell for an implicit RANS iteration}
\begin{tabular}{|p{6cm}|c|c|}
\hline
\textbf{Component} & \textbf{Estimated FLOPs/Cell} & \textbf{Percentage of Total (Approx.)} \\
\hline
Residual calculation (fluxes, turbulence) & \textasciitilde$6\hspace{1pt}\text{k}$ & 14\% \\
\hline
Jacobian matrix assembly & \textasciitilde$12\hspace{1pt}\text{k}$ & 27\% \\
\hline
Linear system solution ($N_{\text{GMRES}}=10$) & \textasciitilde$26.2\hspace{1pt}\text{k}$ & 59\% \\
\hline
Total & \textasciitilde$44.2\hspace{1pt}\text{k}$ & 100\% \\
\hline
\end{tabular}
\end{table}

\subsection{Cost analysis of a wall-modeled LES explicit solver}
Explicit LES requires time-accurate simulation, for which explicit methods are well-suited. A multi-stage Runge-Kutta (RK) scheme is common. The total work per time step is simply $s$ (the number of stages) times the cost of a single spatial operator (residual) evaluation, $R(w)$. No linear system is solved.
The spatial operator $R(w)$ consists of three main parts:
\begin{enumerate}
    \item Convective fluxes (JST scheme): This part is dominated by the artificial dissipation (a blend of second- and fourth-order terms) needed for stability, costing \textasciitilde800 FLOPs per cell per stage.
    \item SGS Model (WALE): The subgrid-scale model amounts to \textasciitilde300 FLOPs per cell per stage.
    \item Flux evaluation: The central difference fluxes and viscous terms add another \textasciitilde300 FLOPs per cell per stage.
\end{enumerate}
The cost of the wall model is confined to boundary cells and is considered negligible in the domain-averaged cost.

\subsubsection{Synthesis and total FLOP count for LES}
The cost of a single residual evaluation is the sum of its components (\textasciitilde$1.4\hspace{1pt}\text{k}$ FLOPs per cell per stage). For a typical two-stage Runge-Kutta scheme ($s=2$), the total cost per time step is:

\begin{equation}
\text{Total cost} = 2 \times 1.4 \hspace{1pt}\text{k} \approx 2.8\hspace{1pt}\text{k} \text{FLOPs per cell per time-step} \ .
\end{equation}

\begin{table}[h!]
\centering
\caption{Estimated FLOPs breakdown per cell for an explicit WMLES time step.}
\begin{tabular}{|p{3cm}|c|c|p{2cm}|c|p{2cm}|c|}
\hline
Component & FLOPs per cell per stage & Total FLOPs per cell (2-stage RK) & Percentage of total (approx.) \\
\hline
Flux evaluation (central diff. + viscous) & \textasciitilde$300$ & \textasciitilde$600$ & $21\%$ \\
\hline
Artificial dissipation (JST) & \textasciitilde$800$ & \textasciitilde $1.6\hspace{1pt}\text{k}$ & $57\%$ \\
\hline
SGS Model (WALE) & \textasciitilde$300$ & \textasciitilde$600$ & $21\%$ \\
\hline
Total & $\textasciitilde1.4\hspace{1pt}\text{k}$ & $\textasciitilde2.8\hspace{1pt}\text{k}$ & $100\%$ \\
\hline
\end{tabular}
\end{table}

\subsection{FLOPs to time-to-solution}
To translate total FLOPs into wall-clock time, one must use the hardware's sustained, not peak, performance. CFD codes are memory-bandwidth-bound, so a sustained efficiency factor, $\eta_{sustained}$, must be used:

\begin{equation}
P_{\text{sustained}} = P_{\text{peak}} \times \eta_{\text{sustained}} \ .
\end{equation}

For stencil-based and sparse matrix codes, a realistic $\eta_{\text{sustained}}$ is typically between 5\% to 15\%. We assume a lower value of 5\% for our low-fidelity implicit RANS solver and a higher level of 15\% for our explicit WMLES solver.

\subsection{Final caveats}
This work provides a first-order estimate. There are several key uncertainties:
\begin{itemize}
    \item The sustained efficiency factor ($\eta_{\text{sustained}}$): This is the largest source of uncertainty, depending on code implementation and memory access patterns. 
    \item Algorithmic variations: The use of different numerical schemes, turbulence models, or linear solvers (e.g., a different preconditioner) will alter the $f$ value.
\end{itemize}
\section{Estimating the number of simulations required}
\label{sec:appendixC}
The parameter space for computational fluid dynamics is exceptionally high-dimensional. In the following section we provide a very approximate estimate on the number of samples $N$ required to build a foundational model. We propose that we can split the simulations required into four core areas: 
\begin{enumerate}
    \item Domains $N_{d}$: High-level application areas (e.g., automotive, aerospace, turbo-machinery, combustion).
    \item Archetypes $N_{a}$: Canonical base geometries within each domain (e.g., a ``sedan'' or ``commercial wide-body aircraft'').
    \item Parameters $N_{p}$: Key geometric and flow variables defining the design space (e.g., Reynolds number, angle of attack).
    \item Values per parameter $N_{pv}$: The samples required per parameter e.g a range of inlet velocities or geometry variations.    
\end{enumerate}

We calculate the total number of simulations, $n_r$ through Equation \ref{eq:datasetestimate}. All these values shown in Table \ref{tab:derivation} are approximates and created to provide an order of magnitude estimation and to provoke further more rigorous estimates.

\begin{equation}
\label{eq:datasetestimate}
    N = N_{d} \times N_{a} \times N_{p} \times N_{pv} 
\end{equation}

\begin{table}[h]
  \caption{Derivation total simulation count}
  \label{tab:derivation}
  \centering
  \begin{tabular}{lcccc}
    \toprule
    $N_{d}$ & $N_{a}$ & $N_{p}$ & $N_{pv}$ & $N$  \\
    \midrule
    $9$ & $100$ & $50$ & $50$  & $2.25 \times 10^{6}$ \\
    \bottomrule
  \end{tabular}
\end{table}

\subsection{Ground transportation}
\begin{itemize}
    \item \textbf{Archetypes (100 total):} e.g Passenger cars \& luxury vehicles, commercial \& industrial haulage, public transport \& service vehicles, rail \& tracked transport, motorcycles \& micromobility, human-powered transport, construction \& agricultural machinery, motorsport \& performance, military \& specialized off-road, recreational \& traditional 
    \item \textbf{Key parameters (D=50):} e.g Vehicle geometry \& bodywork, aerodynamic devices \& appendages, tyre wheel \& suspension geometry, powertrain \& propulsion values, thermal \& energy management, off-road \& terrain interaction, fluid flow \& boundary layer physics   
\end{itemize}

\subsection{Aerospace}
\begin{itemize}
    \item \textbf{Archetypes (100 total):} e.g Commercial transport aircraft, general aviation \& special mission aircraft, rotorcraft \& vertical lift, advanced air mobility vehicles, unmanned aerial systems, light sport \& gliding aircraft, spacecraft \& launch systems, lighter-than-air vehicles, amphibious \& ground effect craft, experimental aerodynamic configurations 
    \item \textbf{Key parameters (D=50):} e.g General aircraft geometry, wing \& airfoil design, empennage \& control surfaces, propulsion \& inlet configuration, rotor \& propeller parameters, mass \& balance properties, flight kinematics \& attitude, freestream \& atmospheric conditions, thermodynamic \& heat transfer variables
\end{itemize}

\subsection{Turbo-machinery and energy systems}
\begin{itemize}
    \item \textbf{Archetypes (100 total):} e.g Aircraft propulsion \& gas turbines, steam \& vapor power cycles, hydropower turbines, wind energy converters, industrial compressors \& blowers, ventilation \& air handling, liquid pumps \& pumping systems, marine propulsion \& thrusters, specialized rotormachinery \& components
    \item \textbf{Key parameters (D=50):} e.g Blade \& airfoil geometry, rotor, stator \& volute dimensions, machine sizing \& clearances, operational flow conditions, fluid thermodynamics \& properties, performance \& dimensionless parameters
\end{itemize}

\subsection{Built environment}
\begin{itemize}
    \item \textbf{Archetypes (100 total):} e.g Basic geometric primitives \& shapes, building morphologies \& typologies, urban planning \& street layouts, geographic \& settlement patterns, bridges \& crossings, civil infrastructure \& water works, transportation \& traffic nodes, industrial, energy \& utility structures, residential, commercial \& public venues
    \item \textbf{Key parameters (D=50):} e.g Urban morphology \& density metrics, building geometry \& architectural features, surface roughness \& terrain topography, computational domain \& boundary conditions, fluid dynamic similarity parameters, turbulence \& flow field physics, thermal environment \& solar radiation, material properties \& thermodynamics, atmospheric stability \& air quality 
\end{itemize}

\subsection{Combustion and reactive flows}
\begin{itemize}
    \item \textbf{Archetypes (100 total):} e.g Fundamental flame types \& configurations, industrial burners \& combustion systems, furnaces, kilns \& boilers, internal combustion engine (ICE) chambers, aerospace \& gas turbine combustors, rocket propulsion \& spacecraft thrusters, fire safety \& hazard dynamics, wildfire \& environmental fire types, atomization \& droplet dynamics 
    \item \textbf{Key parameters (D=50):} e.g Combustor \& burner geometry, flow field \& turbulence scales, stoichiometry \& thermodynamics, fuel physical properties, chemical kinetics \& flame speed, combustion dimensionless groups, radiation \& emissions physics 
\end{itemize}

\subsection{Marine \& hydrodynamics}
\begin{itemize}
    \item \textbf{Archetypes (100 total):} e.g Container ships, bulk carriers, tankers, passenger vessels, offshore energy units, service \& workboats, fishing \& aquaculture vessels, leisure \& recreational vessels
    \item \textbf{Key parameters (D=50):} Principal hull dimensions, hull form coefficients \& features, hydrostatics \& mass properties, propeller geometry, propulsion performance factors, rudder geometry, environmental conditions, fluid properties
\end{itemize}

\subsection{Biological and life sciences}
\begin{itemize}
    \item \textbf{Archetypes (100 total):} e.g Cardiac chambers \& anatomy, heart valves \& prosthetics, vascular system \& pathologies, respiratory airways \& lungs, general organ systems, mechanical circulatory support \& filtration, vascular omplants \& stents, medical delivery \& access systems, microfluidic mixing \& droplet dynamics etc
    \item \textbf{Key parameters (D=50):} e.g anatomical \& geometric dimensions, medical device specifications, tissue mechanics \& surface properties, systemic hemodynamic parameters, fluid dynamic similarity \& shear metrics, biofluid rheology \& viscosity, mass transport \& gas exchange, respiratory \& pulmonary variables, microfluidic \& interfacial physics, computational boundaries \& lumped parameters etc
\end{itemize}

\subsection{Chemical \& process engineering}
\begin{itemize}
    \item \textbf{Archetypes (100 total):} e.g Chemical \& biological reactors, mixing \& homogenization qquipment, centrifuges \& cyclonic separators, filtration \& gravity settling systems, mass transfer columns \& internals, heat exchangers \& reboilers, drying \& evaporative cooling systems, valves \& flow control, pumps \& fluid machinery, piping components \& fittings
    \item \textbf{Key parameters (D=50):} e.g mixing \& vessel geometry, separation \& column internals, heat Exchanger design specs, fluid dynamic similarity parameters, reaction kinetics \& energetics, heat \& mass transfer rates, multiphase flow \& particle physics, process performance \& timescales
\end{itemize}

\subsection{Electronics cooling \& thermal management}
\begin{itemize}
    \item \textbf{Archetypes (100 total):} e.g heatsinks \& extended surfaces, phase change \& heat spreaders, heatsinks \& extended surfaces, liquid cooling components, fans \& air movers, semiconductor packaging, PCB \& discrete components, power supply \& energy storage, server \& rack architecture, data center thermal infrastructure, system chassis \& enclosures
    \item \textbf{Key parameters (D=50):} e.g Semiconductor \& package dimensions, heatsink \& extended surface geometry, PCB \& thermal interface specifications, airflow \& fan geometry, liquid cooling \& micro-channel specs, thermal loads \& operating temperatures, thermal resistance \& impedance metrics, fan performance \& flow dynamics, material properties \& thermophysics, heat transfer coefficients \& similarity numbers
\end{itemize}

\clearpage

\section{Python code for scaling plots}
\label{sec:code}
\begin{lstlisting}[language=Python]

import numpy as np, matplotlib.pyplot as plt

plt.rcParams.update({'font.size':22,'axes.labelsize':22,'axes.titlesize':22,'xtick.labelsize':22,'ytick.labelsize':22,'legend.fontsize':16,'figure.figsize':(10,6),'axes.grid':True,'grid.alpha':0.3,'lines.linewidth':5.5})

# Constants & Pre-calculations
N, S, FX = np.linspace(0, 2e6, 100), 1e6, '#1f77b4'
K_DG = 8.0 / (0.15 * 8e13 * 3600)
K_MT = 8.0 / (0.6 * 5e15 * 3600) * 5.2
SZ = (5e8 / 64) * 1024 * 6

# Cost Functions
def c_dg(n, f=2800, c=5e8, t=2e5): return n * c * t * f * K_DG
def c_mt(n, a=.24, b=.43, e=100):  return n * 50 * e * (SZ + 32**3 * (n**(b/a))) * K_MT

# Base Costs
yd, yt = c_dg(N), c_mt(N)

# Plot Configurations: (Name, Title, Range, CostFn, LabelFn, FixedY, FixedLbl, FixedLS, FixedLW, VaryLS)
cfgs = [
    ('alpha', r'$\alpha$', np.linspace(.22,.26,5), lambda v: c_mt(N,a=v,b=.425), lambda v: fr'Training cost ($\alpha$={v:.3f})', yd, 'Data generation cost', '--', 2.5, '-'),
    ('beta', r'$\beta$', np.linspace(.4,.47,5), lambda v: c_mt(N,b=v), lambda v: fr'Training cost ($\beta$={v:.3f})', yd, 'Data generation cost', '--', 2.5, '-'),
    ('epochs', 'Epochs', [10,50,100,200,500], lambda v: c_mt(N,e=v), lambda v: f'Training cost (Epochs={int(v)})', yd, 'Data generation cost', '--', 2.5, '-'),
    ('datagen_flops', 'FLOPS/cell/step', np.linspace(2e3,1e4,5), lambda v: c_dg(N,f=v), lambda v: f'Data gen cost ({int(v)} FLOPS)', yt, 'Training cost (Fixed)', '-', 3, '--'),
    ('cells', 'cells', np.linspace(1e8,1e9,5), lambda v: c_dg(N,c=v), lambda v: f'Data gen cost ({v/1e9:.1f}B cells)' if v>=1e9 else f'Data gen cost ({int(v/S)}M cells)', yt, 'Training cost (Fixed)', '-', 3, '--'),
    ('timesteps', 'timesteps', np.linspace(5e4,5e5,5), lambda v: c_dg(N,t=v), lambda v: f'Data gen cost ({int(v/1000)}k Steps)', yt, 'Training cost (Fixed)', '-', 3, '--')
]

for nm, tit, vals, fn, lbl, fy, fl, fls, flw, vls in cfgs:
    f, ax = plt.subplots()
    ax.plot(N/S, fy/S, label=fl, color=FX, ls=fls, lw=flw)
    for v, c in zip(vals, plt.cm.magma(np.linspace(0.2, 0.8, len(vals)))):
        ax.plot(N/S, fn(v)/S, label=lbl(v), color=c, linewidth=2.5, ls=vls)
    ax.set(xlabel='Sample size (millions)', ylabel='Cost ($ millions)', title=f'Cost vs. sample size (varying {tit})', yscale='log', ylim=(1,1000), xlim=(0,2))
    ax.legend(loc='lower right'); f.tight_layout(); f.savefig(f'cost_analysis_{nm}.png', dpi=300)

plt.show()
\end{lstlisting}


\end{document}